\documentclass[a4paper,usenames,dvipsnames,11pt]{article}
\pdfoutput=1

\usepackage{jheppub}
\usepackage{slashed}
\usepackage{mathrsfs,booktabs,multirow,tabularx}
\usepackage{stmaryrd}
\usepackage{xspace}
\usepackage{fancyvrb}
\usepackage[makeroom]{cancel}
\usepackage{amsmath}    
\usepackage{amssymb}    %
\usepackage{graphicx}   
\usepackage{verbatim}   
\usepackage{lscape}
\usepackage{subfig}
\usepackage{listings}
\usepackage{mathtools}

\def\beq{\begin{equation}}
\def\beqn{\begin{eqnarray}}
\def\eeq{\end{equation}}
\def\eeqn{\end{eqnarray}}

\def\MSbeq#1{eq.~({\bf II}.#1)}
\def\ePDFteq#1{eq.~({\bf II}.#1)}

\def\ePDF#1{\Gamma_{\!#1}}

\def\binomial#1#2{
\left(\!\!
\begin{array}{c}
#1\\
#2
\end{array}
\!\!\right)
}
\def\poch#1#2{\left(#1\right)_{#2}}
\DeclarePairedDelimiter\ceil{\lceil}{\rceil}
\DeclarePairedDelimiter\floor{\lfloor}{\rfloor}

\newcommand\sss{\scriptscriptstyle}

\newcommand\half{\frac{1}{2}}
\newcommand\as{\alpha_{\sss S}}
\newcommand\aem{\alpha}

\newcommand\Daem{\Delta\aem}
\newcommand\aemz{\alpha(\mu_0)}

\newcommand\aemmu{\alpha(\mu)}

\newcommand\gE{\gamma_{\sss\rm E}}

\newcommand{\bN}{\bar{N}}

\newcommand{\epem}{e^+e^-}
\newcommand{\lp}{e^+}
\newcommand{\lm}{e^-}

\newcommand{\ord}{{\cal O}}

\newcommand\NF{n_{\sss F}}

\newcommand\hsig{\hat{\sigma}}

\newcommand\hS{\hat{S}}

\newcommand\APmat{{\mathbb P}}
\newcommand\Eop{{\mathbb E}}
\newcommand\Mmat{{\mathbb M}}
\newcommand\Kmat{{\mathbb K}}

\newcommand\MSb{\overline{\rm MS}}

\newcommand\muz{\mu_0}

\newcommand\dencpar{M_1}
\newcommand\dendpar{M_2}

\newcommand\Fzero{F_0}
\newcommand\Fone{F_1}
\newcommand\Ftwo{F_2}

\newcommand{\Melleq}{\stackrel{\infty}{=}}
\newcommand\stepf{\Theta}
\newcommand\xiDt{\xi_2}
\newcommand\hxiDt{\hat{\xi}_2}
\newcommand\invm{\mathfrak{M}}

\title{On factorisation schemes for the electron parton distribution
functions in QED}

\author[a,b]{S. Frixione,}
\affiliation[a]{INFN, Sezione di Genova, Via Dodecaneso 33, I-16146, 
Genoa, Italy}
\affiliation[b]{ PH Department, TH Division, CERN, CH-1211 Geneva 23, 
Switzerland}

\emailAdd{Stefano.Frixione@cern.ch}

\abstract{
The electron, positron, and photon Parton Distribution Functions 
(PDFs) of the unpolarised electron have recently been computed at the 
next-to-leading logarithmic accuracy in QED, by adopting the
$\overline{\rm MS}$ factorisation scheme. We present here analogous
results, obtained by working in a different framework that is inspired
by the so-called DIS scheme. We derive analytical solutions relevant
to the large-$z$ region, where we show that the behaviour of the PDFs
depends in a dramatic way on whether running-$\alpha$ effects are
included to all orders, as opposed to being truncated to some fixed order.
By means of suitable initial and evolution conditions, next-to-leading 
logarithmic accurate PDFs are obtained whose large-$z$ functional forms
are identical to those of their leading logarithmic counterparts.
}


\preprint{
\begin{flushright}
CERN-TH-2021-078\\
\end{flushright}
}

\begin{document}
\maketitle
\flushbottom

\section{Introduction\label{sec:intro}}
The computation of $\epem$ cross sections in perturbative QED leads
to the emergence of logarithms of the ratio $m^2/E^2$, where $m$ is
the electron mass and $E$ a scale of the order of the hardness
of the process, such as the collider energy. Owing to the smallness
of the electron mass, these logarithms are numerically very large 
in virtually all situations of phenomenological interest, in particular
at the future colliders presently under consideration. This is problematic,
since it implies that the perturbative series is not well behaved,
with coefficients that grow with the power of $\aem$. It is therefore
mandatory to re-sum such logarithms in order to obtain meaningful physical 
predictions. While there are different sources of large logarithms, possibly
dependent on the observables one considers, for some classes of them,
associated with soft and/or collinear radiation off initial- and final-state
particles, general resummation techniques exist, such as 
YFS~\cite{Yennie:1961ad,Jadach:2000ir}, parton shower~\cite{Anlauf:1991wr,
Fujimoto:1993ge,Munehisa:1995si,CarloniCalame:2000pz}, and factorisation 
formulae~\cite{Kuraev:1985hb,Ellis:1986jba} (the latter sometimes 
improperly referred to as structure function approach).

Factorisation formulae are fully analogous to their QCD counterparts and,
as in that case, resum large logarithms of collinear initial- and final-state
origin by means of parton distribution functions (PDFs) and fragmentation 
functions, respectively. At variance with what happens in QCD, these 
quantities are computable in perturbation theory. As far as the PDFs are
concerned, which are the focus of this paper, results of leading-logarithmic
(LL) accuracy~\cite{Skrzypek:1990qs,Skrzypek:1992vk,Cacciari:1992pz} have
recently been extended to the next-to-leading logarithmic (NLL) 
accuracy~\cite{Frixione:2019lga,Bertone:2019hks}\footnote{For an 
approach that uses the language of renormalisation group equations,
see e.g.~refs.~\cite{Blumlein:2007kx,Blumlein:2011mi}.}. Among other things,
this extension has a strong phenomenological motivation, in view of
the precision targets relevant to future colliders.

At the NLL and beyond, PDFs depend on the choice of the factorisation scheme.
In essence, this is the choice of the finite part in the subtraction of the 
remaining (after the cancellations stemming from the definition of IRC-safe
observables) initial-state collinear singularities, the residue of whose 
poles in $1/\bar{\epsilon}$ is an Altarelli-Parisi kernel. Such a finite
part enters both the short-distance cross sections and the PDFs so that,
when both are expanded in $\aem$ at the same order, it cancels exactly.
However, since in practical applications the PDFs must not be expanded
(because otherwise they lose the capability of resumming the logarithms),
a dependence on the scheme choice is present, although beyond the 
accuracy of the computation.

While the factorisation-scheme dependence is suppressed by powers
of $\aem$ w.r.t.~terms which are under formal control, one may wonder 
whether numerically this suppression is effective, i.e.~whether 
the coefficients that multiply these $\aem^k$ factors can grow
pathologically large. Naively, the $\MSb$ result of ref.~\cite{Bertone:2019hks}
would seem to imply that this is the case. In the $z\to 1$ region,
which gives the dominant contribution to the cross section, the
electron and photon densities (inside an electron) read as 
follow\footnote{While the definitions of all of the quantities
that enter eqs.~(\ref{NLLsol3run}) and~(\ref{ePDFgaasy}) will be given
later (see eqs.~(\ref{tdef}), (\ref{xiR1def}), (\ref{chi1Rdef}),
(\ref{dores}), and~(\ref{dtres})), this is unimportant in this generic 
discussion, where the only relevant matter is the functional form in $z$.
The latter is apparent, knowing that typical parameters values are 
$\xi_1\simeq 0.05$, $\hat{\xi}_1\simeq 0.04$, $A(\xi_1)\simeq 19.8$, 
$B(\xi_1)\simeq -1.5$, and $t\simeq 0.025$.}:
\beqn
\!\!\!\!\!\Gamma_{\lm}(z,\mu)&\stackrel{z\to 1}{\longrightarrow}&
\frac{e^{-\gE\xi_1}e^{\hat{\xi}_1}}{\Gamma(1+\xi_1)}\,
\xi_1(1-z)^{-1+\xi_1}
\label{NLLsol3run}
\\*&&\phantom{aaa}\times
\Bigg\{1+\frac{\aem(\mu_0)}{\pi}\Bigg[\left(\log\frac{\mu_0^2}{m^2}-1\right)
\left(A(\xi_1)+\frac{3}{4}\right)-2B(\xi_1)+\frac{7}{4}
\nonumber\\*&&\phantom{\times aaa1+\frac{\aem}{\pi}\Bigg[}\;
+\left(\log\frac{\mu_0^2}{m^2}-1-2A(\xi_1)\right)\log(1-z)
-\log^2(1-z)\Bigg]\Bigg\}\,,
\nonumber
\\*
\ePDF{\gamma}(z,\mu)&\stackrel{z\to 1}{\longrightarrow}&
\frac{t\,\aemz^2}{\aemmu}\,\frac{3}{2\pi\xi_1}\,\log(1-z)
-\frac{t\,\aemz^3}{\aemmu}\,\frac{1}{2\pi^2\xi_1}\,\log^3(1-z)\,,
\label{ePDFgaasy}
\eeqn
whereas their LL counterparts do not feature any of the 
\mbox{$\log(1-z)$} terms that appear on the r.h.s.~of eqs.~(\ref{NLLsol3run})
and~(\ref{ePDFgaasy}). However, this is indeed naive. Firstly, the PDFs
are unphysical objects: the actual impact of these logarithms must
be assessed after the convolution with the cross sections. Secondly,
the leading behaviour of the electron PDF is in any case given by the 
prefactor (i.e.~the first $z$-dependent term on the r.h.s.~of 
eq.~(\ref{NLLsol3run})), that appears at the LL as well and that has 
an integrable singularity much stronger than logarithmic.

However, the question of the impact of the residual scheme dependence on 
physical observables remains relevant. Since such a dependence is not
parametrical, the best way to assess it is that of comparing results
obtained in different factorisation schemes; while not completely conclusive,
this allows one to see in practice at which level of accuracy the scheme
dependence may constitute a phenomenological problem.

In this paper we shall compute the NLL-accurate electron PDFs
by adopting a factorisation scheme alternative to the $\MSb$ employed
in ref.~\cite{Bertone:2019hks}; in particular, we shall use a definition
whose physical motivation is the same as the one that underpins the
so-called DIS scheme in QCD~\cite{Altarelli:1978id}. At variance with
ref.~\cite{Bertone:2019hks}, where analytical results have been presented
for both the $z\to 1$ region (to all orders in $\aem$) and the small-
and intermediate-$z$ region (up to $\ord(\aem^3$)), in this paper we shall
limit ourselves to studying analytically the $z\to 1$ regime since, as is 
shown in eqs.~(\ref{NLLsol3run}) and~(\ref{ePDFgaasy}), this is where the
dominant effects are expected to appear; a comprehensive
phenomenological study, that includes the numerical implementation of the 
results presented here, will be given elsewhere~\cite{BCFSZZ}.

This paper is organised as follows. In sect.~\ref{sec:eopK} we introduce
the operator that will be used to solve the PDF evolution equations,
and do so in a generic factorisation scheme. The scheme that is actually 
employed is called $\Delta$, and is defined in sect.~\ref{sec:Del}.
In sect.~\ref{sec:NSasy} we derive the asymptotic $z\to 1$ form
of the non-singlet PDF in the $\Delta$ scheme by solving its evolution
equation in two different ways, namely by retaining running-$\aem$
effects to all orders (sect.~\ref{sec:asyD1}), or by truncating them at the 
first non-trivial order in $\aem$ (sect.~\ref{sec:asyD2}). The comparison
between these two results is discussed in sect.~\ref{sec:D1vsD2}. In 
sect.~\ref{sec:Sgaasy} we apply the same strategy as in sect.~\ref{sec:asyD1}
to the singlet-photon sector. Finally, in sect.~\ref{sec:concl} we present
our conclusions. Some Mellin-transform results that are used throughout
the paper are collected in appendix~\ref{sec:Mell}.

\section{Evolution operator\label{sec:eopK}}
In this section we write the evolution equations for the 
PDFs~\cite{Gribov:1972ri,Lipatov:1974qm,Altarelli:1977zs,Dokshitzer:1977sg}
in the same way as it has been done in ref.~\cite{Bertone:2019hks}, namely
by introducing a PDF evolution operator, and by defining it so as it 
works in a generic factorisation scheme. In the interest of a shorter
notation, throughout this paper eq.~(x.y) of ref.~\cite{Bertone:2019hks} 
will be denoted by \MSbeq{x.y}. The procedure we follow here has significant
overlaps with standard works in QCD (e.g.~refs.~\cite{Furmanski:1981cw,
Diemoz:1987xu}); however, a peculiarity of the QED case warrants a 
compact re-derivation.

The evolution equations for a given particle type\footnote{That is, 
an electron, a photon, and so forth; in this paper we shall consider 
explicitly only the case of an electron, but the treatment presented 
here is general. The reader is encouraged to check sect.~2 of 
ref.~\cite{Frixione:2019lga} for the definitions of particles and
partons used in the present context.} are written in Mellin space as 
in \MSbeq{4.6}:
\beq
\frac{\partial \Gamma_N(\mu)}{\partial\log\mu^2}=
\frac{\aem(\mu)}{2\pi}\,\APmat_N(\mu)\,\Gamma_N(\mu)=
\sum_{k=0}^\infty\left(\frac{\aem(\mu)}{2\pi}\right)^{k+1}
\APmat_N^{[k]}\,\Gamma_N(\mu)\,.
\label{matAPmell}
\eeq
Here, we have denoted by $\Gamma_N$ a column vector that collects
all of the relevant PDFs:
\beq
\Gamma_N=\left(
\begin{array}{c}
\ePDF{\alpha_1,N}\\
\ePDF{\alpha_2,N}\\
\vdots\\
\ePDF{\alpha_n,N}\\
\end{array}
\right)\,,
\label{FN}
\eeq
with $\ePDF{\alpha_i}$ the PDF of parton $\alpha_i$ inside the
particle of interest. Conventionally, we shall label the partons
so that $\alpha_1$ coincides with the identity of the particle (thus,
for an electron particle, $\alpha_1=\lm$). The PDFs are regarded as
evolved from the values they assume at a given (small) scale $\muz$
(initial conditions) by means of an evolution operator (an $n\times n$
matrix), thus:
\beq
\Gamma_N(\mu)=\Eop_N(\mu,\muz)\,\Gamma_{0,N}(\muz)\,,\;\;\;\;\;\;
\Eop_N(\muz,\muz)=I\,.
\label{FvsE}
\eeq
Equation~(\ref{matAPmell}) is then fully equivalent to:
\beqn
\frac{\partial \Eop_N(\mu,\muz)}{\partial\log\mu^2}&=&
\sum_{k=0}^\infty\left(\frac{\aem(\mu)}{2\pi}\right)^{k+1}
\APmat_N^{[k]}\,\Eop_N(\mu,\muz)
\nonumber\\*
&=&
\frac{\aem(\mu)}{2\pi}\left[\APmat_N^{[0]}+
\frac{\aem(\mu)}{2\pi}\,\APmat_N^{[1]}\right]\Eop_N(\mu,\muz)
+\ord(\aem^2)\,.
\label{matAPmell3}
\eeqn
More conveniently, in order to take the running of $\aem$ into account
in an easier manner, one uses the variable $t$ of \MSbeq{4.10} instead of 
the scale $\mu$:
\beq
t=\frac{1}{2\pi b_0}\log\frac{\aem(\mu)}{\aem(\mu_0)}\,,
\label{tdef}
\eeq
to re-express eq.~(\ref{matAPmell3}) as follows:
\beqn
\frac{\partial \Eop_N(t)}{\partial t}&=&
\frac{b_0\aem^2(\mu)}{\beta(\aem(\mu))}
\sum_{k=0}^\infty\left(\frac{\aem(\mu)}{2\pi}\right)^k
\APmat_N^{[k]}\,\Eop_N(t)
\nonumber\\*
&=&
\left[\APmat_N^{[0]}+\frac{\aem(\mu)}{2\pi}\left(
\APmat_N^{[1]}-\frac{2\pi b_1}{b_0}\,\APmat_N^{[0]}\right)
\right]\Eop_N(t)+\ord(\aem^2)\,.
\label{matAPmell4}
\eeqn
In order to simplify the upcoming discussion, we consider the convolution
of the PDFs with the short-distance cross sections for only one of the two
incoming partons; it should be clear that this entails no loss of
generality. Collecting the subtracted partonic cross section again
into a column vector
\beq
\hS=\left(
\begin{array}{c}
\hsig_{\alpha_1}\\
\vdots\\
\hsig_{\alpha_n}\\
\end{array}
\right)\,,
\eeq
each element of which thus corresponds to a partonic channel, the
particle-level cross section in Mellin space reads as follows:
\beq
\sigma_N(\mu)=\hS_N^{\rm T}(\mu)\,\Gamma_{N+1}(\mu)=
\hS_N^{\rm T}(\mu)\,\Eop_{N+1}(\mu,\muz)\,\Gamma_{0,N+1}(\muz)\,.
\label{physxsec}
\eeq
The RGE invariance of the cross section is therefore:
\beq
\frac{\partial\sigma_N(\mu)}{\partial\log\mu^2}=0+\ord(\aem^{b+k+1})\,,
\eeq
where $b$ is the power of $\alpha$ that factors out at the Born level,
and $k$ is the accuracy of the computation ($k=0$ being the LO, $k=1$ the 
NLO, and so forth -- in practice, in what follows we shall understand the 
factor $\alpha^b$, and restrict ourselves to the NLO case $k=1$, thus
writing $\ord(\aem^2)$ to denote terms beyond NLO accuracy).

So far, we have implicitly assumed to work in the $\MSb$ factorisation scheme; 
this is in keeping with what has been done in ref.~\cite{Bertone:2019hks},
and applies in particular to the RGEs of the evolution operator.
In order to employ a different scheme, we start from the partonic 
cross sections. In the FKS formalism~\cite{Frixione:1995ms,
Frixione:1997np}\footnote{Needless to say, the result presented here
do not depend on the subtraction formalism adopted. However, it is useful,
in order to be definite, to make a specific choice.} the arbitrariness
of the definition of the factorisation scheme is parametrised in terms
of a set of generalised functions $K_{ij}(z)$ that enter the so-called
$(n+1)$-body degenerate cross sections, whose contributions to an NLO
partonic cross section can be written as follows in the Mellin space
(see e.g.~eqs.~(3.21)--(3.23) of ref.~\cite{Frederix:2018nkq}):
\beq
\hsig_{{\alpha_i},N}^{(K)}=\ldots-\frac{\aem(\mu)}{2\pi}\sum_{j=1}^n
K_{{\alpha_j\alpha_i},N+1}\hsig_{{\alpha_j},N}^{[0]}\,.
\label{sigik}
\eeq
Here, we have included an index ``$K$'' to make it explicit the fact
that, in general, we work in a factorisation scheme different from
$\MSb$; we point out that these formulae still encompass the $\MSb$
case, which corresponds to choosing $K_{ij}(z)\equiv 0$. By  collecting
the $K_{ij}(z)$ functions into an $n\times n$ matrix:
\beq
\Kmat=\left(
\begin{array}{ccc}
K_{\alpha_1\alpha_1} & \ldots & K_{\alpha_1\alpha_n} \\
\vdots & \ddots & \vdots \\
K_{\alpha_n\alpha_1} & \ldots & K_{\alpha_n\alpha_n} \\
\end{array}
\right)\,,
\label{Kmat}
\eeq
the matrix form of eq.~(\ref{sigik}) reads:
\beq
\hS_{N}^{(K)}(\mu)=\hS_N(\mu)-\frac{\aem(\mu)}{2\pi}\,
\Kmat_{N+1}^{\rm T}\,\hS_N^{[0]}(\mu)
=\left(I-\frac{\aem(\mu)}{2\pi}\,
\Kmat_{N+1}^{\rm T}\right)\hS_N(\mu)
+\ord(\aem^2)\,.
\label{SKN}
\eeq
As the rightmost side explicitly indicates, the two forms of $\hS_{N}^{(K)}$
that appear in eq.~(\ref{SKN}) are strictly equivalent at the NLO.

The elementary nature of the asymptotic states in QED renders it possible
to compute perturbatively the initial conditions for PDF evolution.
The NLO results of ref.~\cite{Frixione:2019lga} for such initial conditions
are given in a generic factorisation scheme, and feature the same
functions $K_{ij}(z)$ that appear above (for the electron particle,
see in particular eqs.~(4.121) and~(4.189) there). The scheme-dependence
of the PDF initial conditions is a beyond-LO effect, and therefore at
the LO we can write (by keeping in mind the conventions established
in eq.~(\ref{FN})):
\beq
\Gamma_{0,N}^{[0]}\equiv \Gamma_{0,N}^{(K)[0]}=\left(
\begin{array}{c}
1\\
0\\
\vdots\\
0\\
\end{array}
\right)\,.
\label{F0N}
\eeq
It is then easy to see that the NLO results of ref.~\cite{Frixione:2019lga} 
can be written as follows:
\beqn
\Gamma_{0,N}^{(K)}(\muz)&=&\Gamma_{0,N}(\muz)+\frac{\aem(\mu_0)}{2\pi}\,
\Kmat_N\,\Gamma_{0,N}^{[0]}
\label{FK0N}
\\*
&=&\left(I+\frac{\aem(\mu_0)}{2\pi}\,\Kmat_N\right)\Gamma_{0,N}(\muz)
+\ord(\aem^2)\,,
\label{FK0N3}
\eeqn
where the rightmost side follows from the same considerations as those
employed in eq.~(\ref{SKN}). We point out that, owing to eq.~(\ref{F0N}),
at the NLO only the first column of the $\Kmat$ matrix gives a non-null
contribution; in other words, only the $K_{\alpha_i\alpha_j}$ elements with
$\alpha_j$ equal to the particle identity (i.e.~$\alpha_1$) are relevant. 
This is in keeping with the elementary nature of that particle, so that
at the NLO it is the branchings \mbox{$\alpha_1\to\alpha_i+X$} that
are sufficient to control the factorisation scheme. Thus, at this level
of precision, the choice of the $K_{\alpha_i\alpha_j}$ elements with
$\alpha_j\ne\alpha_1$ is essentially arbitrary, since they are associated
with NNLO contributions. Note that this is not true for the partonic
short distance cross sections, where those functions may contribute at
the NLO as well; this is however irrelevant as far as PDF evolution is
concerned, which is our goal here.

In a generic scheme, we write the analogue of eq.~(\ref{FvsE})
with a minimal change of notation, namely:
\beq
\Gamma_{N}^{(K)}(\mu)=\Eop_{N}^{(K)}(\mu,\mu_0)\,\Gamma_{0,N}^{(K)}(\muz)
\,,\;\;\;\;\;\;\Eop_{N}^{(K)}(\mu_0,\mu_0)=I\,,
\label{FvsEK}
\eeq
from which the analogue of eq.~(\ref{physxsec}) follows:
\beq
\sigma_{N}^{(K)}(\mu)=\hS_{N}^{(K)\rm T}(\mu)\,
\Eop_{N+1}^{(K)}(\mu,\muz)\,\Gamma_{0,N+1}^{(K)}(\muz)\,.
\label{Kphysxsec}
\eeq
The factorisation-scheme independence of any physical observable
is then equivalent to requiring that:
\beq
\sigma_N=\sigma_{N}^{(K)}+\ord(\aem^2)\,.
\label{sNeqsKN}
\eeq
For the l.h.s.~of this equation we shall use eq.~(\ref{physxsec}).
Conversely, for its r.h.s.~we can replace eqs.~(\ref{SKN}) and~(\ref{FK0N3})
into eq.~(\ref{FvsEK}):
\beq
\sigma_{N}^{(K)}=
\hS_N^{\rm T}
\left(I-\frac{\aem(\mu)}{2\pi}\,\Kmat_{N+1}\right)
\Eop_{N+1}^{(K)}
\left(I+\frac{\aem(\mu_0)}{2\pi}\,\Kmat_{N+1}\right)\Gamma_{0,N+1}\,.
\label{Kphysxsec2}
\eeq
Thus, by replacing eqs.~(\ref{physxsec}) and~(\ref{Kphysxsec2}) into
eq.~(\ref{sNeqsKN}), and by canceling the factors that appear on both sides
(i.e.~the partonic cross sections and the PDF initial conditions),
we arrive at:
\beq
\Eop_N=
\left(I-\frac{\aem(\mu)}{2\pi}\,\Kmat_N\right)
\Eop_{N}^{(K)}
\left(I+\frac{\aem(\mu_0)}{2\pi}\,\Kmat_N\right)\,.
\label{EopvsEopK0}
\eeq
In keeping with the fact that it stems from eq.~(\ref{sNeqsKN}),
eq.~(\ref{EopvsEopK0}) is understood to hold up to $\ord(\aem^2)$
terms. In fact, at $\mu=\muz$ we obtain, from the initial conditions
in eqs.~(\ref{FvsE}) and~(\ref{FvsEK}):
\beq
I=I-\left(\frac{\aem(\mu_0)}{2\pi}\right)^2\Kmat_N^2\,.
\label{IeqI}
\eeq
Although this is within the accuracy at which we are working, it also
suggests that the identity of eq.~(\ref{EopvsEopK0}) can be conveniently 
modified by observing that, for perturbation theory to work, there must 
exist a non-zero measure set in the space of the Mellin variable where the 
following inverse operator exists and is well defined:
\beq
\left(I+\frac{\aem(\mu)}{2\pi}\,\Kmat_N\right)^{-1}=
I+\sum_{k=1}^{\infty}(-)^k\left(\frac{\aem(\mu)}{2\pi}\right)^k 
\Kmat_N^k\,.
\label{invIpJ}
\eeq
Since the first two terms in the series on the r.h.s.~of eq.~(\ref{invIpJ})
coincide with the terms within the leftmost brackets in
eq.~(\ref{EopvsEopK0}), it follows that at $\ord(\aem^2)$ we can 
also write:
\beq
\Eop_N=
\left(I+\frac{\aem(\mu)}{2\pi}\,\Kmat_N\right)^{-1}
\Eop_{N}^{(K)}
\left(I+\frac{\aem(\mu_0)}{2\pi}\,\Kmat_N\right)\,,
\label{EopvsEopK}
\eeq
which we shall consider as valid to all orders in $\aem$ (note that
the analogue of eq.~(\ref{IeqI}) in this case would read \mbox{$I=I$}).
The sought evolution equation for the $\Eop_{N}^{(K)}$ operator can be
obtained by deriving both sides of eq.~(\ref{EopvsEopK}) w.r.t.~the
variable $t$, and by exploiting eq.~(\ref{matAPmell4}). In order to
do that, we observe that eq.~(\ref{invIpJ}) implies:
\beq
\frac{\partial}{\partial t}
\left(I+\frac{\aem(\mu)}{2\pi}\,\Kmat_N\right)^{-1}=
-b_0\aem(\mu)\Kmat_N
\left(I+\frac{\aem(\mu)}{2\pi}\,\Kmat_N\right)^{-2}\,,
\label{Eaemder}
\eeq
from whence:
\beqn
\frac{\partial \Eop_{N}^{(K)}(t)}{\partial t}&=&
b_0\aem(\mu)\Kmat_N
\left(I+\frac{\aem(\mu)}{2\pi}\,\Kmat_N\right)^{-1}
\Eop_{N}^{(K)}(t)
\label{EKopev}
\\*&+&
\frac{b_0\aem^2(\mu)}{\beta(\aem(\mu))}
\sum_{k=0}^\infty\left(\frac{\aem(\mu)}{2\pi}\right)^k
\nonumber
\\*&&\phantom{\frac{b_0\aem^2(\mu)}{\beta(\aem(\mu))}}\times
\left(I+\frac{\aem(\mu)}{2\pi}\,\Kmat_N\right)\APmat_N^{[k]}
\left(I+\frac{\aem(\mu)}{2\pi}\,\Kmat_N\right)^{-1}\Eop_{N}^{(K)}(t)\,.
\nonumber
\eeqn
This result generalises the $\MSb$ one in the first line of 
eq.~(\ref{matAPmell4}) and, as it should, coincides with it when 
$\Kmat=0$. An $\ord(\aem^2)$-accurate expansion then leads to:
\beqn
\frac{\partial \Eop_{N}^{(K)}(t)}{\partial t}&=&
\left[\APmat_N^{[0]}+\frac{\aem(\mu)}{2\pi}\left(
\APmat_N^{[1]}-\frac{2\pi b_1}{b_0}\,\APmat_N^{[0]}\right)
\right]\Eop_{N}^{(K)}(t)
\nonumber\\*&+&
\frac{\aem(\mu)}{2\pi}\left(
\left[\Kmat_N,\APmat_N^{[0]}\right]
+2\pi b_0 \Kmat_N\right)\Eop_{N}^{(K)}(t)\,,
\label{EKopev2}
\eeqn
which generalises the second line of eq.~(\ref{matAPmell4}) and coincides 
with it when $\Kmat=0$. As consistency demands, eq.~(\ref{EKopev2}) can also 
be obtained directly from eq.~(\ref{EopvsEopK0}), by deriving w.r.t.~the 
variable $t$ the two sides of that equation.

The solution of the RGE for the evolution operator that stems from
eq.~(\ref{EKopev2}) is the analogue of what is usually done in
QCD (see e.g.~refs.~\cite{Diemoz:1987xu,Cacciari:1998it}), whereas
to the best of our knowledge the QCD counterpart of eq.~(\ref{EKopev})
has never been considered. We shall show in the following that this is
justified: while the difference between these two equations is of NNLO in 
both QCD and in QED, in the latter case it leads to dramatically different 
behaviours in the $z\to 1$ region; this does not happen in QCD.

\section{A DIS-like factorisation scheme\label{sec:Del}}
The RGEs for the evolution operator, eqs.~(\ref{EKopev}) and~(\ref{EKopev2}),
are valid in any factorisation scheme; here, we shall make a definite
choice for one such scheme.

The electron PDF initial conditions at the NLO~\cite{Frixione:2019lga},
namely:
\beqn
\ePDF{i}^{[0]}(z,\muz)&=&\delta_{i\lm}\delta(1-z)\,,
\label{G0sol}
\\
\ePDF{\lm}^{[1]}(z,\muz)&=&\left[\frac{1+z^2}{1-z}\left(
\log\frac{\muz^2}{m^2}-2\log(1-z)-1\right)\right]_+ +K_{ee}(z)\,,
\label{G1sol2}
\\
\ePDF{\gamma}^{[1]}(z,\muz)&=&\frac{1+(1-z)^2}{z}\left(
\log\frac{\muz^2}{m^2}-2\log z-1\right) +K_{\gamma e}(z)\,,
\label{Ggesol2}
\\
\ePDF{\lp}^{[1]}(z,\muz)&=&0\,,
\label{Gpossol2}
\eeqn
show that it is straightforward to choose the $K_{ij}(z)$ functions
so that the $\ord(\aem)$ contributions at $\muz=m$ vanish, thus making
the NLO initial conditions to coincide with their LO counterparts. This
condition is reminiscent of the definition of the DIS scheme in
QCD~\cite{Altarelli:1978id}; in fact, in view of the stricter similarity
between the case of the electron PDFs in QED and that of the $b$-quark 
perturbative fragmentation function in QCD~\cite{Mele:1990cw}, we shall
call the scheme thus defined as $\Delta$ scheme, for consistency with
the notation of ref.~\cite{Cacciari:1998it}. From eqs.~(\ref{G1sol2})
and~(\ref{Ggesol2}) we thus define:
\beqn
K_{ee}^{(\Delta)}(z)&=&\left[\frac{1+z^2}{1-z}
\Big(2\log(1-z)+1\Big)\right]_+\,,
\label{Kdelz}
\\
K_{\gamma e}^{(\Delta)}(z)&=&
\frac{1+(1-z)^2}{z}\Big(2\log z+1\Big)\,.
\label{Kgadelz}
\eeqn
We point out that eqs.~(\ref{Kdelz}) and~(\ref{Kgadelz}), in addition
to fulfilling the requirement above, are also consistent with the 
momentum-conservation condition for the branchings of an electron
(see eq.~(5.5) of ref.~\cite{Frixione:2019lga}).

The factorisation-scheme matrix $\Kmat$ of eq.~(\ref{Kmat}) is then
defined as follows\footnote{Similarly to what was done in
refs.~\cite{Frixione:2019lga,Bertone:2019hks}, in practical applications
we limit ourselves to considering the electron, the photon, and the
positron.}:
\beq
\Kmat^{(\Delta)}=\left(
\begin{array}{ccc}
K_{ee}^{(\Delta)}(z) & 0 & 0 \\
K_{\gamma e}^{(\Delta)}(z) & 0 & K_{\gamma e}^{(\Delta)}(z) \\
0 & 0 & K_{ee}^{(\Delta)}(z) \\
\end{array}
\right)\,.
\label{KmatDel}
\eeq
As was discussed in sect.~\ref{sec:eopK}, the matrix elements of $\Kmat$ whose 
column index is not equal to the identity of the particle of interest (the 
electron here, first column) give beyond-NLO effects. By setting them equal 
to zero as in the second column of eq.~(\ref{KmatDel}) (that corresponds to the 
photon), we impose conditions that are identical to what one had to enforce 
if one were to define a DIS-like scheme for the {\em photon} particle, per 
eqs.~(5.42) and~(5.47) of ref.~\cite{Frixione:2019lga}. As far as the
positron is concerned (third column), its matrix elements are motivated
by charge-invariance considerations. Although the positron plays an 
essentially trivial role here, the form of eq.~(\ref{KmatDel}) leads to 
simple analytical computations, in particular to the immediate decoupling 
of the evolution of the singlet and non-singlet components.

We shall employ eq.~(\ref{KmatDel}) when solving the RGE for the evolution
operator. For the latter, we shall use both eq.~(\ref{EKopev}) and
eq.~(\ref{EKopev2}), and distinguish the solutions thus obtained by
referring to the former as $\Delta_1$, and to the latter as $\Delta_2$.

\section{Non-singlet large-$z$ solutions\label{sec:NSasy}}
As is customary, solving the RGE of the evolution operator is easier
if one uses the non-singlet (NS), singlet (S), and photon PDFs, since 
the non-singlet decouples and thus one works in a one-dimensional flavour
space in that case, which leads to scalar (as opposed to matrix) equations.
The reader can find all of the relevant definitions for the evolution
in terms of NS, S, and $\gamma$ in sect.~3 of ref.~\cite{Bertone:2019hks}.

\subsection{$\Delta_1$ result\label{sec:asyD1}}
According to the definition given at the end of sect.~\ref{sec:Del},
the $\Delta_1$ solution is that obtained by solving eq.~(\ref{EKopev}) 
in the $\Delta$ scheme. For the NS case, the only relevant scheme-changing
function is that of eq.~(\ref{Kdelz}). A direct computation leads to
the following Mellin-space result:
\beq
K_{ee,N}^{(\Delta)}\equiv K_N^{(\Delta)}\,\Melleq\,
2\Big(\log^2\bN-\log\bN+C\Big)\,,
\label{KNasy}
\eeq
with:
\beq
C=\frac{\pi^2}{6}-1
\label{Cconst}
\eeq
and
\beq
\bN=N\,e^{\gE}\,.
\label{bNdef}
\eeq
In eq.~(\ref{KNasy}) we have introduced the symbol $\Melleq$ in order
to understand that terms which are subleading\footnote{How much subleading
depends on the context: while for the NS solution all terms suppressed by
powers of $1/N$ are discarded, the $1/N$ ones (i.e.~the first subleading)
will need to be kept in the S-$\gamma$ sector, as detailed in
sect.~\ref{sec:Sgaasy}.} in $N$ in the $N\to\infty$ limit are discarded.
The same symbol will be employed in the case of the inverse Mellin transform,
to discard terms subleading when $z\to 1$.

Equation eq.~(\ref{KNasy}) is then employed in the solution of
eq.~(\ref{EKopev}), for which we adopt a notation more in keeping with
the scalar nature of the NS case, by means of the following formal
replacements:
\beq
\Eop_{N}^{(K)}\;\longrightarrow\;E_{N}^{(\Delta_1)}\,,
\;\;\;\;\;\;\;\;
\APmat_N^{[k]}\;\longrightarrow\;P_N^{[k]}\,,
\;\;\;\;\;\;\;\;
\Kmat_N\;\longrightarrow\;K_N^{(\Delta)}\,.
\label{NSrepl}
\eeq
By using eq.~(\ref{tdef}), it is straightforward to see that
eq.~(\ref{EKopev}) in the NS case has the following solution
(note that with eq.~(\ref{KNasy}) the inverse operator of 
eq.~(\ref{invIpJ}) exists not only in the asymptotic region
$N\to\infty$, which would be sufficient, but everywhere in the
range of interest, \mbox{$N\in [1,\infty)$}. This is convenient 
for the matching of the analytical large-$z$ solution with its
numerical counterpart, relevant to smaller $z$'s):
\beq
\log E_{N}^{(\Delta_1)}=
\log E_N+\log\left(\frac{1+\frac{\aem(\mu)}{2\pi}K_N^{(\Delta)}}
{1+\frac{\aem(\mu_0)}{2\pi}K_N^{(\Delta)}}\right)\,,
\label{Esol3}
\eeq
where the first term on the r.h.s.~is the $\MSb$ solution
of \MSbeq{5.50}
\beq
\log E_N\,\Melleq\,
-\xi_1\log\bN+\hat{\xi}_1\,,
\label{ENxi1}
\eeq
with the $\xi_1$ and $\hat{\xi}_1$ parameters defined 
in \MSbeq{5.51} and \MSbeq{5.53}: 
\beqn
\xi_1&=&
2t-\frac{\aem(\mu)}{4\pi^2 b_0}\Big(1-e^{-2\pi b_0 t}\Big)
\left(\frac{20}{9}\NF+\frac{4\pi b_1}{b_0}\right)\,,
\label{xiR1def}
\\
\hat{\xi}_1&=&\frac{3}{2}\,t+\frac{\aem(\mu)}{4\pi^2 b_0}
\Big(1-e^{-2\pi b_0 t}\Big)
\left(\lambda_1-\frac{3\pi b_1}{b_0}\right)\,.
\label{chi1Rdef}
\eeqn
Here, $b_0$ and $b_1$ are the coefficients of the QED $\beta$ function:
\beqn
&&\frac{\partial\aem(\mu)}{\partial\log\mu^2}=\beta(\aem)=
b_0\aem^2+b_1\aem^3+\ldots\,,
\label{betaQED}
\\
&&b_0=\frac{\NF}{3\pi}\,,\;\;\;\;\;\;\;\;
b_1=\frac{\NF}{4\pi^2}\,,
\label{b0b1}
\eeqn
and we have defined:
\beq
\lambda_1=\frac{3}{8}-\frac{\pi^2}{2}+6\zeta_3-\frac{\NF}{18}(3+4\pi^2)\,.
\label{lambda1}
\eeq
With eqs.~(\ref{FvsEK}) 
and~(\ref{G1sol2}), and by bearing in mind that the term in the latter 
equation that multiplies the \mbox{$\log(\muz^2/m^2)$} factor is the 
lowest-order Altarelli-Parisi kernel in the NS sector, we obtain the 
initial condition in the $\Delta$ scheme in the Mellin space:
\beq
\Gamma_{0,N}^{(\Delta)}=1+\frac{\aem(\mu_0)}{2\pi}L_0P_N^{[0]}\,,
\label{inicDelta}
\eeq
where we have introduced the shorthand notation:
\beq
L_0=\log\frac{\mu_0^2}{m^2}\,,
\eeq
and the sought large-$N$ solution for the NS PDF:
\beq
\Gamma_N^{(\Delta_1)}\,\Melleq\,
\left(1+\frac{\aem(\mu_0)}{2\pi}L_0P_N^{[0]}\right)
e^{-\gE\xi_1}e^{\hat{\xi}_1}N^{-\xi_1}\,
\frac{1+\frac{\aem(\mu)}{2\pi}K_N^{(\Delta)}}
{1+\frac{\aem(\mu_0)}{2\pi}K_N^{(\Delta)}}\,.
\label{tmp10}
\eeq
For consistency with eq.~(\ref{KNasy}), we shall use
\beq
P_N^{[0]}\,\Melleq\,
2\left(-\log\bN+\frac{3}{4}\right)
\label{P0Nasy}
\eeq
in this section. We are interested in determining the $z$-space form
of the NS PDF in eq.~(\ref{tmp10}), which implies computing its inverse
Mellin transform. In order to proceed, we re-express its rightmost
term as a series in $\aem$, thus:
\beqn
\Gamma_N^{(\Delta_1)}&\Melleq&
\left(1+\frac{\aem(\mu_0)}{2\pi}L_0P_N^{[0]}\right)
e^{-\gE\xi_1}e^{\hat{\xi}_1}N^{-\xi_1}
\nonumber\\*&\times&
\left(1+\frac{\aem(\mu)-\aem(\mu_0)}{\pi}\sum_{j=1}^\infty
(-)^{j-1}\left(\frac{\aem(\mu_0)}{\pi}\right)^{j-1}
\left(\frac{K_N^{(\Delta)}}{2}\right)^j\right)\,.
\label{tmp11}
\eeqn
With eq.~(\ref{KNasy}), we see that the scheme-changing terms in
eq.~(\ref{tmp11}) are the following polynomials:
\beq
\left(\frac{K_N^{(\Delta)}}{2}\right)^j\,\Melleq\,
\Big(\log^2\bN-\log\bN+C\Big)^j=
\sum_{k=0}^{2j}L_{j,k}\log^{2j-k}\bN\,,
\label{Ktoj}
\eeq
where, after some algebra\footnote{It is understood that if the lower entry 
of a binomial coefficient is larger than the upper one that coefficient
is set equal to zero.\label{ft:binom}}:
\beq
L_{j,k}=(-)^k\sum_{q=0}^{\floor*{k/2}}
(-)^q\,\frac{\poch{q-k}{q}}{q!}\binomial{j}{k-q}C^{\,q}\,,
\label{Lcoeff2}
\eeq
which is expressed in terms of the Pochhammer symbol:
\beq
\poch{x}{n}=\frac{\Gamma(x+n)}{\Gamma(x)}=x(1+x)\ldots (n-1+x)\,.
\label{pochdef}
\eeq
By using the algebraic identities above, eq.~(\ref{tmp11}) can be recast
as follows:
\beq
\Gamma_N^{(\Delta_1)}\,\Melleq\,
e^{-\gE\xi_1}e^{\hat{\xi}_1}N^{-\xi_1}
\left[\left(1+\frac{3\aem(\mu_0)}{4\pi}L_0\right)S_1
-\frac{\aem(\mu_0)}{\pi}L_0S_2\right],
\label{fKN}
\eeq
where:
\beqn
S_1\!&=&\!1+\frac{\aem(\mu)-\aem(\mu_0)}{\pi}\sum_{j=1}^\infty
(-)^{j-1}\left(\frac{\aem(\mu_0)}{\pi}\right)^{j-1}
\sum_{k=0}^{2j}L_{j,k}\log^{2j-k}\bN\,,\phantom{aaa}
\label{S1def}
\\
S_2\!&=&\!S_1\log\bN\,.
\label{S2def}
\eeqn
By using the result of eq.~(\ref{invMlogqN}), we can express the
inverse Mellin transforms relevant to eq.~(\ref{fKN}) as follows:
\beqn
&&M^{-1}\!\left[N^{-\xi_1}\sum_{k=0}^{2j}\,
L_{j,k}\log^{2j-k}\bN\right]\,\Melleq\,
\frac{\xi_1(1-z)^{-1+\xi_1}}{\Gamma(1+\xi_1)}
\label{invMS1el}
\\&&\phantom{aaaaaaaaaaa}\times
\sum_{p=0}^{2j}\left[\sum_{k=0}^p (-)^p\binomial{2j-k}{p-k} L_{j,k}\,
d_{p-k}(\xi_1)\right]\log^{2j-p}(1-z)\,,
\nonumber\\
&&\!\!\!\!M^{-1}\!\left[N^{-\xi_1}\log\bN\,
\sum_{k=0}^{2j}L_{j,k}\log^{2j-k}\bN\right]\,\Melleq\,
\frac{\xi_1(1-z)^{-1+\xi_1}}{\Gamma(1+\xi_1)}
\label{invMS2el}
\\&&\phantom{aaaa}\times
\sum_{p=0}^{2j+1}\left[\sum_{k=0}^p (-)^{p-1}\binomial{2j+1-k}{p-k} 
L_{j,k}\,d_{p-k}(\xi_1)\right]\log^{2j+1-p}(1-z)\,.
\nonumber
\eeqn
In eqs.~(\ref{invMS1el}) and~(\ref{invMS2el}) the index $p$ labels
the subleading-ness of the corresponding \mbox{$\log(1-z)$} term: 
the larger $p$, the more subleading the associated contribution. 
The idea is that of computing the inverse Mellin transforms of 
\mbox{$N^{-\xi_1}S_1$} and \mbox{$N^{-\xi_1}S_2$} 
by employing eqs.~(\ref{invMS1el}) and~(\ref{invMS2el}) after having 
exchanged the order of the summations over $j$ and $p$.
Thus:
\beqn
M^{-1}\!\left[N^{-\xi_1}S_1\right]&\Melleq&
\frac{\xi_1(1-z)^{-1+\xi_1}}{\Gamma(1+\xi_1)}
\sum_{p=0}^{\infty} {\cal S}_{1,p}(z)\,,
\label{invMS1}
\\
M^{-1}\!\left[N^{-\xi_1}S_2\right]&\Melleq&
\frac{\xi_1(1-z)^{-1+\xi_1}}{\Gamma(1+\xi_1)}
\sum_{p=0}^{\infty} {\cal S}_{2,p}(z)\,,
\label{invMS2}
\eeqn
with:
\beqn
{\cal S}_{1,p}(z)&=&\delta_{p0}\,\frac{\aem(\mu)}{\aem(\mu_0)}
+\frac{\aem(\mu)-\aem(\mu_0)}{\pi}\,
\frac{\hat{{\cal S}}_{1,p}(z)}{D(z)^{p+1}}\,,
\label{cS10def}
\\
{\cal S}_{2,p}(z)&=&\frac{\aem(\mu)}{\aem(\mu_0)}
\Big(-\delta_{p0}\log(1-z)+\delta_{p1}d_1(\xi_1)\Big)
+\frac{\aem(\mu)-\aem(\mu_0)}{\pi}\,
\frac{\hat{{\cal S}}_{2,p}(z)}{D(z)^{p+1}}\,,\phantom{aaaa}
\label{cS20def}
\eeqn
where, from eqs.~(\ref{invMS1el}) and~(\ref{invMS2el}):
\beqn
\frac{\hat{{\cal S}}_{1,p}(z)}{D(z)^{p+1}}&=&
-\frac{\pi\delta_{p0}}{\aem(\muz)}+
(-)^p\sum_{j=\ceil*{p/2}}^\infty\left(1-\delta_{j0}\right)
(-)^{j-1}\left(\frac{\aem(\muz)}{\pi}\right)^{j-1}
\label{S1presf}
\\*&&\phantom{aaaaaa}\times
\sum_{k=0}^{p}\binomial{2j-k}{p-k}L_{j,k}\,
d_{p-k}(\xi_1)\log^{2j-p}(1-z)\,,
\nonumber
\\
\frac{\hat{{\cal S}}_{2,p}(z)}{D(z)^{p+1}}&=&
\frac{\pi}{\aem(\muz)}
\Big(\delta_{p0}\log(1-z)-\delta_{p1}d_1(\xi_1)\Big)
\label{S2presf}
\\*&&
-(-)^p\!\!\!\sum_{j=\ceil*{(p-1)/2}}^\infty\!\!\left(1-\delta_{j0}\right)
(-)^{j-1}\left(\frac{\aem(\muz)}{\pi}\right)^{j-1}
\nonumber
\\*&&\phantom{aaaaaaaa}\times
\sum_{k=0}^{p}\binomial{2j+1-k}{p-k}L_{j,k}\,
d_{p-k}(\xi_1)\log^{2j+1-p}(1-z)\,.
\nonumber
\eeqn
We have defined\footnote{Note that eqs.~(\ref{S1presf})
and~(\ref{S2presf}) define the functions \mbox{$\hat{{\cal S}}_{i,p}(z)$},
in a way motivated by the fact that the factors \mbox{$D(z)^{-p-1}$} are 
seen to emerge from the summations on the r.h.s.'s. Furthermore, the
terms proportional to $\delta_{p0}$ and $\delta_{p1}$ are such that the 
functions \mbox{${\cal S}_{1,0}(z)$}, \mbox{${\cal S}_{2,0}(z)$}, and
\mbox{${\cal S}_{2,1}(z)$} in eqs.~(\ref{cS10def}) and~(\ref{cS20def})
include the contributions stemming from the first terms on the r.h.s.~of
eqs.~(\ref{S1def}) and~(\ref{S2def}), respectively.}:
\beq
D(z)=1+\frac{\aem(\mu_0)}{\pi}\,\log^2(1-z)\,,
\label{Ddendef}
\eeq
and the ${\cal S}_{i,p}$ functions are such that:
\beqn
\frac{\hat{{\cal S}}_{1,p}(z)}{D(z)^{p+1}}
&\stackrel{z\to 1}{\longrightarrow}&\log^{-2-p}(1-z)\;
\longrightarrow\;0\,,
\label{S1psupp}
\\
\frac{\hat{{\cal S}}_{2,p}(z)}{D(z)^{p+1}}
&\stackrel{z\to 1}{\longrightarrow}&\log^{-1-p}(1-z)\;
\longrightarrow\;0\,,
\label{S2psupp}
\eeqn
which show again the role of the index $p$ in determining the 
subleading-ness of a given contribution. In practice, this implies
keeping only the first few terms in the sums over $p$ in eqs.~(\ref{invMS1})
and~(\ref{invMS2}), for which one can compute explicitly the 
${\cal S}_{i,p}$ functions, by carrying out the summations over $j$
that enter their definitions. The results for the lowest $p$ values are:
\beqn
&&\hat{{\cal S}}_{1,0}(z)=-\frac{\pi}{\aem(\mu_0)}\,,
\label{hS10}
\\
&&\hat{{\cal S}}_{1,1}(z)=\big(1-2d_1(\xi_1)\big)\log(1-z)\,,
\label{hS11}
\\
&&\hat{{\cal S}}_{1,2}(z)=\frac{\aem(\mu_0)}{\pi}\left(C-1
+3d_1(\xi_1)-3d_2(\xi_1)\right)\log^2(1-z)
\nonumber
\\*&&\phantom{\hat{{\cal S}}_{1,2}(z)=}
+C-d_1(\xi_1)+d_2(\xi_1)\,,
\label{hS12}
\\*
&&\hat{{\cal S}}_{2,0}(z)=\frac{\pi}{\aem(\mu_0)}\log(1-z)\,,
\label{hS20}
\\
&&\hat{{\cal S}}_{2,1}(z)=-\big(1-d_1(\xi_1)\big)\log^2(1-z)
-\frac{\pi}{\aem(\mu_0)}\,d_1(\xi_1)\,,
\label{hS21}
\\
&&\hat{{\cal S}}_{2,2}(z)=-\frac{\aem(\mu_0)}{\pi}\left(C-1
+2d_1(\xi_1)-d_2(\xi_1)\right)\log^3(1-z)
\nonumber
\\*&&\phantom{\hat{{\cal S}}_{2,2}(z)=}
-\Big(C-2d_1(\xi_1)+3d_2(\xi_1)\Big)\log(1-z)\,.
\label{hS22}
\eeqn
In view of a numerical implementation,
we also have computed these contributions up to $p=5$, which are too long to
be reported here; the extension to yet larger $p$ values poses no problem, 
but it is not particularly useful in view of eqs.~(\ref{S1psupp}) 
and~(\ref{S2psupp}).

In summary, the NLL-accurate large-$z$ expression of the NS PDF in the 
$\Delta$ scheme obtained by solving eq.~(\ref{EKopev}) for the RGE of 
the evolution operator is:
\beqn
\Gamma_{\rm\sss NS}^{(\Delta_1)}(z,\mu)&\stackrel{z\to 1}{\longrightarrow}&
\frac{e^{-\gE\xi_1}e^{\hat{\xi}_1}}{\Gamma(1+\xi_1)}\,
\xi_1(1-z)^{-1+\xi_1}
\label{NLLsol8Del}
\\*&&\times
\left[\left(1+\frac{3\aem(\mu_0)}{4\pi}L_0\right)
\sum_{p=0}^{\infty} {\cal S}_{1,p}(z)
-\frac{\aem(\mu_0)}{\pi}L_0\sum_{p=0}^{\infty} {\cal S}_{2,p}(z)\right].
\nonumber
\eeqn
This implies that the strict asymptotic form of eq.~(\ref{NLLsol8Del}),
where all terms that vanish at $z\to 1$ are ignored, reads as follows
(by taking eq.~(\ref{dores}) into account):
\beqn
\Gamma_{\rm\sss NS}^{(\Delta_1)}(z,\mu)\!&\stackrel{z\to 1}{\longrightarrow}&\!
\frac{e^{-\gE\xi_1}e^{\hat{\xi}_1}}{\Gamma(1+\xi_1)}\,
\xi_1(1-z)^{-1+\xi_1}
\nonumber\\*&&\phantom{aa}\times
\left[\frac{\aem(\mu)}{\aem(\mu_0)}+
\frac{\aem(\mu)}{\pi}L_0
\left(A(\xi_1)+\log(1-z)+\frac{3}{4}\right)\right].\phantom{aaaa}
\label{NLLsol8Delasy}
\eeqn
The comparison of eq.~(\ref{NLLsol8Delasy}) with the solution\footnote{At
the NLO, the $z\to 1$ expressions of the electron and the non-singlet PDFs
coincide.} of eq.~(\ref{NLLsol3run}) shows that all \mbox{$\log(1-z)$} terms 
that appear in the latter\footnote{In addition to those one would obtain
by expanding the \mbox{$(1-z)^{-1+\xi_1}$} prefactor. Except for
checks on its perturbative behaviour~\cite{Bertone:2019hks}, such an
expansion must never be carried out.} (bar for that multiplied by $L_0$;
note that usually one sets $L_0=0$) are features of the $\MSb$ 
scheme, and disappear if the $\Delta$ scheme is employed. We also
point out that eq.~(\ref{NLLsol8Delasy}) could have been directly obtained
from eq.~(\ref{tmp10}), by taking the strict $N\to\infty$ limit of the 
scheme-dependent contribution to the evolution operator:
\beq
\frac{1+\frac{\aem(\mu)}{2\pi}K_N^{(\Delta)}}
{1+\frac{\aem(\mu_0)}{2\pi}K_N^{(\Delta)}}
\;\longrightarrow\;
\frac{\aem(\mu)}{\aem(\mu_0)}\,,
\eeq
and by employing eq.~(\ref{invMlogqN}) with $q=0,1$.

\subsection{$\Delta_2$ result\label{sec:asyD2}}
In this section we shall again obtain the asymptotic form of the NS
PDF in the $\Delta$ scheme; at variance with the procedure of
sect.~\ref{sec:asyD1}, here this will be done by employing eq.~(\ref{EKopev2})
rather than eq.~(\ref{EKopev}).
This leads to the following solution for the evolution operator, which
is the analogue of eq.~(\ref{Esol3}):
\beqn
\log E_{N}^{(\Delta_2)}&\Melleq&-\xi_1\log\bN+\hat{\xi}_1
+\Daem\,\frac{K_N^{(\Delta)}}{2}
\nonumber
\\*
&\Melleq&-\xi_1\log\bN+\hat{\xi}_1
+\Daem\,\Big(\log^2\bN-\log\bN+C\Big)\,,
\label{EKNxi1}
\eeqn
where we have used eqs.~(\ref{KNasy}) and~(\ref{ENxi1}), and defined:
\beq
\Daem=\frac{\aem(\mu)-\aem(\mu_0)}{\pi}\,.
\label{Daemdef}
\eeq
Equation~(\ref{EKNxi1}) suggests to introduce the following parameters:
\beqn
\xiDt&=&\xi_1+\Daem\,,
\label{xi23mxi1}
\\
\hxiDt&=&\hat{\xi}_1+C\,\Daem\,,
\label{chi3mchi1}
\eeqn
so that:
\beq
\log E_{N}^{(\Delta_2)}\,\Melleq\,-\xiDt\log\bN+\hxiDt+
\Daem\,\log^2\bN\,.
\label{Esol2Delta3}
\eeq
With the initial condition in the $\Delta$ scheme, eq.~(\ref{inicDelta}),
we therefore arrive at the analogue of eq.~(\ref{tmp10}), namely:
\beq
\Gamma_N^{(\Delta_2)}\,\Melleq\,
\left(1+\frac{\aem(\mu_0)}{2\pi}L_0P_N^{[0]}\right)
e^{-\gE\xiDt}e^{\hxiDt}N^{-\xiDt}\,
\exp\Big[\Daem\,\log^2\bN\Big]\,.
\label{tmp12C}
\eeq
The computation of the inverse Mellin transform of eq.~(\ref{tmp12C})
proceeds similarly to what was done in sect.~\ref{sec:asyD1}. One
first expands the rightmost exponent in eq.~(\ref{tmp12C}), thus obtaining
the analogue of eq.~(\ref{tmp11}):
\beq
\Gamma_N^{(\Delta_2)}\,\Melleq\,
\left(1+\frac{\aem(\mu_0)}{2\pi}L_0P_N^{[0]}\right)
e^{-\gE\xiDt}e^{\hxiDt}N^{-\xiDt}
\sum_{j=0}^\infty\frac{\Daem^j}{j!}\,\log^{2j}\bN\,.
\label{tmp13}
\eeq
From eq.~(\ref{tmp13}) we arrive at the analogue of eq.~(\ref{fKN}):
\beq
\Gamma_N^{(\Delta_2)}\,\Melleq\,
e^{-\gE\xiDt}e^{\hxiDt}N^{-\xiDt}
\left[\left(1+\frac{3\aem(\mu_0)}{4\pi}L_0\right)T_1
-\frac{\aem(\mu_0)}{\pi}L_0T_2\right],
\label{fKN2}
\eeq
where:
\beqn
T_1\!&=&\!\sum_{j=0}^\infty\frac{\Daem^j}{j!}\,\log^{2j}\bN\,,\phantom{aaa}
\label{Tf1def}
\\
T_2\!&=&\!T_1\log\bN\,.
\label{Tf2def}
\eeqn
In the computation of the inverse Mellin transform of eq.~(\ref{fKN2}) 
one can exploit eqs.~(\ref{invMS1el}) and~(\ref{invMS2el}). In this way, 
we arrive at the analogues of eqs.~(\ref{invMS1}) and~(\ref{invMS2}), 
namely:
\beqn
M^{-1}\!\left[N^{-\xiDt}T_1\right]&\Melleq&
\frac{\xiDt(1-z)^{-1+\xiDt}}{\Gamma(1+\xiDt)}\,
e^{\Daem\log^2(1-z)}\,
\sum_{p=0}^{\infty} {\cal T}_{1,p}(z)\,,
\label{invMT1}
\\
M^{-1}\!\left[N^{-\xiDt}T_2\right]&\Melleq&
\frac{\xiDt(1-z)^{-1+\xiDt}}{\Gamma(1+\xiDt)}\,
e^{\Daem\log^2(1-z)}\,
\sum_{p=0}^{\infty} {\cal T}_{2,p}(z)\,.
\label{invMT2}
\eeqn
The summands on the r.h.s.~of eqs.~(\ref{invMT1}) and~(\ref{invMT2}) 
have remarkably compact expressions; by construction\footnote{The definitions 
of the \mbox{${\cal T}_{i,p}(z)$} functions in eqs.~(\ref{T1presf}) 
and~(\ref{T2presf}) stem from the fact that the exponential factors on the 
l.h.s.~of those equations emerge naturally from the summations on the r.h.s..}:
\beqn
e^{\Daem\log^2(1-z)}{\cal T}_{1,p}(z)\!&=&\!
d_p(\xiDt)\,(-)^p\!\!
\sum_{j=\ceil*{p/2}}^\infty
\frac{\Daem^j}{j!}\binomial{2j}{p}
\log^{2j-p}(1-z)\,,
\label{T1presf}
\\
e^{\Daem\log^2(1-z)}{\cal T}_{2,p}(z)\!&=&\!
-{d_p(\xiDt)}\,(-)^p\!\!\!\!\!\!
\sum_{j=\ceil*{(p-1)/2}}^\infty\!\!
\frac{\Daem^j}{j!}\binomial{2j+1}{p}
\log^{2j+1-p}(1-z)\,.\phantom{aaa}
\label{T2presf}
\eeqn
The key feature of eqs.~(\ref{invMT1}) and~(\ref{invMT2}) is
the fastly-growing exponential prefactor. Furthermore, contrary to 
what happens in eqs.~(\ref{invMS1}) and~(\ref{invMS2}), terms with 
increasingly large values of $p$'s are not progressively more suppressed 
at $z\to 1$ (although they are increasingly suppressed by powers of $\Daem$).
This implies that, at variance with the case of the $\Delta_1$ solution
of sect.~\ref{sec:asyD1}, the series of $p$ on the r.h.s.~of
eqs.~(\ref{invMT1}) and~(\ref{invMT2}) must be summed in order to obtain 
a sensible result; fortunately, this can be done. By using the explicit 
expressions on the r.h.s.~of eqs.~(\ref{T1presf}) and~(\ref{T2presf}),
after some algebra one arrives at the following results:
\beqn
\sum_{p=0}^\infty {\cal T}_{1,p}(z)&=&
\sum_{k=0}^\infty \frac{\Daem^k}{k!}{\cal T}_{1;k}(z)\,,
\label{T1psum}
\\*
\sum_{p=0}^\infty {\cal T}_{2,p}(z)&=&
\sum_{k=0}^\infty \frac{\Daem^k}{k!}{\cal T}_{2;k}(z)
+\sum_{p=0}^\infty \delta_{1,{\rm mod}(p,2)}\,
\frac{\Daem^{\floor*{p/2}}}{\floor*{p/2}!}\,
d_p(\xiDt)\,,
\label{T2psum}
\eeqn
where the functions ${\cal T}_{i;k}(z)$ are:
\beqn
{\cal T}_{1;k}(z)&=&
\sum_{p=0}^\infty \stepf\!\left(k\le\floor*{p/2}\right)\,d_p(\xiDt)\,
\frac{\left(-2\Daem\log(1-z)\right)^{p-2k}}{(p-2k)!}\,,
\label{T1pkA}
\\*
{\cal T}_{2;k}(z)&=&
-\sum_{p=0}^\infty \stepf\!\left(k\le\floor*{p/2}\right)\,d_p(\xiDt)\,
\frac{p+1}{p+1-2k}
\nonumber\\*&&\phantom{-\sum_{p=0}^\infty}\times
\log(1-z)\,\frac{\left(-2\Daem\log(1-z)\right)^{p-2k}}{(p-2k)!}\,.
\label{T2pkA}
\eeqn
At this point, one can exploit the explicit results for the $d_p$ 
coefficients, given in terms of their generating functional
(eq.~(\ref{dffbygen})). This leads to:
\beqn
{\cal T}_{1;k}(z)&=&
\frac{1}{{\cal G}_d(\xiDt,0)}
\sum_{p=2k}^\infty\,
\left.\frac{\partial^p{\cal G}_d(\xiDt,\delta)}{\partial\delta^p}
\right|_{\delta=0}\!\!
\frac{\left(-2\Daem\log(1-z)\right)^{p-2k}}{(p-2k)!}\,,\phantom{aaa}
\label{T1pkB}
\\*
{\cal T}_{2;k}(z)&=&-\frac{1}{{\cal G}_d(\xiDt,0)}
\sum_{p=2k}^\infty\,
\left.\frac{\partial^p{\cal G}_d(\xiDt,\delta)}{\partial\delta^p}
\right|_{\delta=0}
\frac{p+1}{p+1-2k}
\nonumber\\*&&\phantom{aaaa}\times
\log(1-z)\,\frac{\left(-2\Daem\log(1-z)\right)^{p-2k}}{(p-2k)!}\,,
\label{T2pkB}
\eeqn
with the function ${\cal G}_d$ given in eq.~(\ref{dffgenfun}).
The emergence of Taylor series on the r.h.s.~of eqs.~(\ref{T1pkB}) 
and~(\ref{T2pkB}) is manifest. By assuming the uniform convergence of 
such series, we can re-write those results are follows:
\beqn
{\cal T}_{1;k}(z)&=&
\frac{1}{{\cal G}_d(\xiDt,0)}
\left.\frac{\partial^{2k}{\cal G}_d(\xiDt,\delta)}{\partial\delta^{2k}}
\right|_{\delta=-2\Daem\log(1-z)}\,,
\label{T1pkres22}
\\*
{\cal T}_{2;k}(z)&=&\frac{1}{2\Daem}\frac{1}{{\cal G}_d(\xiDt,0)}
\left.\frac{\partial^{2k}\left[\delta\,{\cal G}_d(\xiDt,\delta)\right]}
{\partial\delta^{2k}}
\right|_{\delta=-2\Daem\log(1-z)}
-\frac{k}{\Daem}\,d_{2k-1}(\xiDt)\,.
\label{T2pkres22}
\eeqn
By replacing these results into eqs.~(\ref{T1psum}) and~(\ref{T2psum}),
respectively, we obtain\footnote{Note that the lowest-order term in
eq.~(\ref{T2psumex}) is of $\ord(\Daem^0)$.}:
\beqn
\sum_{p=0}^\infty {\cal T}_{1,p}(z)&=&
\frac{1}{{\cal G}_d(\xiDt,0)}
\sum_{k=0}^\infty \frac{\Daem^k}{k!}
\left.\frac{\partial^{2k}{\cal G}_d(\xiDt,\delta)}{\partial\delta^{2k}}
\right|_{\delta=-2\Daem\log(1-z)}\,,
\label{T1psumex}
\\*
\sum_{p=0}^\infty {\cal T}_{2,p}(z)&=&
\frac{1}{2\Daem}\frac{1}{{\cal G}_d(\xiDt,0)}
\sum_{k=0}^\infty \frac{\Daem^k}{k!}
\left.\frac{\partial^{2k}\left[\delta\,{\cal G}_d(\xiDt,\delta)\right]}
{\partial\delta^{2k}}\right|_{\delta=-2\Daem\log(1-z)}\,.
\label{T2psumex}
\eeqn
Both sides of eqs.~(\ref{T1psumex}) and~(\ref{T2psumex}) contain summations 
of an infinite number of terms. However, at variance with what happens
on the l.h.s.'s, the summands on the r.h.s.'s have a clear hierarchy,
being suppressed by increasingly large powers of $\Daem$; such a 
hierarchy is not spoiled by the presence of \mbox{$\log(1-z)$} terms, 
which have been resummed in the procedure carried out above.

By putting everything back together, we finally arrive at the analogue
of eq.~(\ref{NLLsol8Del}):
\beqn
&&\Gamma_{\rm\sss NS}^{(\Delta_2)}(z,\mu)
\;\stackrel{z\to 1}{\longrightarrow}\;
\frac{e^{-\gE\xiDt}e^{\hxiDt}}{\Gamma(1+\xiDt)}\,
\xiDt(1-z)^{-1+\xiDt}\,
\frac{e^{\Daem\log^2(1-z)}}{{\cal G}_d(\xiDt,0)}
\sum_{k=0}^\infty \frac{\Daem^k}{k!}
\label{NLLsol9Del}
\\*&&\phantom{aaaaaa}\times
\left.\left[\left(1+\frac{3\aem(\mu_0)}{4\pi}L_0\right)
\frac{\partial^{2k}{\cal G}_d(\xiDt,\delta)}{\partial\delta^{2k}}
-\frac{\aem(\mu_0)}{2\pi\Daem}L_0
\frac{\partial^{2k}\left[\delta\,{\cal G}_d(\xiDt,\delta)\right]}
{\partial\delta^{2k}}\right]\right|_{\delta=-2\Daem\log(1-z)}.
\nonumber
\eeqn
As was the case for the sums over $p$ in eq.~(\ref{NLLsol8Del}),
one can limit oneself to considering only the first few terms in
the sum over $k$ in eq.~(\ref{NLLsol9Del}) to obtain a reliable
numerical prediction.

\subsection{Comments\label{sec:D1vsD2}}
The comparison between the $\Delta_1$ and $\Delta_2$ solutions
given in eqs.~(\ref{NLLsol8Del}) and~(\ref{NLLsol9Del}) shows
dramatic differences in the $z\to 1$ region. While the former
has a fairly smooth behaviour (on top of the integrable-singular
prefactor, which is present also at the LL), the latter features
a non-integrable singularity (induced by the exponential factor
\mbox{$e^{\Daem\log^2(1-z)}$}) with alternating signs (induced
by the function ${\cal G}_d$). It is remarkable that these issues
of the $\Delta_2$ solution stem entirely from the running of $\aem$;
at fixed $\aem$, one has $\Daem=0$, which eliminates the problem.

Ultimately, the difference between the evolution operators relevant
to the $\Delta_1$ and $\Delta_2$ solutions, eqs.~(\ref{Esol3})
and~(\ref{Esol2Delta3}), would be innocuous, were it not for the
fact that the scheme-changing kernel, which is exponentiated, is
driven by a soft-collinear double logarithm, which is the origin 
of the \mbox{$\log^2\bN$} term that is dominant in the Mellin space, 
and that badly spoils the series expansion in $\aem$ of the evolution 
equations at $N\to\infty$; only by keeping all terms in such a series one 
is able to avoid this problem. Lest this fact be misunderstood: we are 
talking here about the logarithmic behaviour of the PDFs (which is what 
is relevant in physical applications). The fixed-order expansions of 
the PDFs (that is not relevant to physics, but useful in QED as a check of
the correctness of the computations) give the expected results:
namely, that eqs.~(\ref{NLLsol8Del}) and~(\ref{NLLsol9Del}) differ by
$\ord(\aem^2)$ terms (i.e.~of NNLO), and that their $\ord(\aem t^0)$ 
coefficient coincides with the initial condition (in the $z\to 1$ limit)
in the $\Delta$ scheme.

We point out that the problem with the $\Delta_2$ solution cannot be studied 
numerically in QED in a direct manner. By setting \mbox{$\aem(\mu)=1/128$}
and \mbox{$\aem(\muz)=1/137$}, the factor \mbox{$e^{\Daem\log^2(1-z)}$}
is equal to one for \mbox{$z=1-10^{-x}$}, with \mbox{$x\simeq 34$},
and no numerical integrator will generate values of $z$ so close to one.
However, the integration of $\epem$ cross sections requires in any
case a change of variables that allows one to efficiently probe
the $z\to 1$ region (which, owing to the presence of the LL-like
integrable-singular prefactor, is relevant to any kind of electron PDF),
and this can only be done by knowing the analytical behaviour of the 
PDF in this region, hence the problem with the $\Delta_2$ solution.

\noindent
$\blacklozenge$ {\em The QCD case.} The evolution operator in QCD obeys
formally the same equations as the QED one, with the obvious formal
exchange $\aem\to\as$, and the crucial difference that the sign in front 
of the $b_0$ term on the second line on the r.h.s.~of eq.~(\ref{EKopev2}) is 
negative rather than positive\footnote{Likewise, the sign in front of the
first terms on the r.h.s.~of eq.~(\ref{EKopev}) must be changed. However,
to the best of our knowledge the analogue of the $\Delta_1$ solution has
not been considered in QCD.}. A simple algebra, that takes into
account the fact that the evolution variables $t$ are defined in
different ways in QCD and QED, shows that in QCD one arrives at a
solution which has the same functional form as that of sect.~\ref{sec:asyD2},
with \mbox{$\Daem\to\Delta\as$}. Needless to say, this applies to
heavy-quark perturbative fragmentation functions\footnote{In addition
to the space- to time-like exchange, final-state evolution involves 
matrix kernels that are the transposed of those relevant to the 
initial state. This does not affect this discussion.}, and not 
to PDFs, since this is the only case where the initial conditions are 
the same as their QED counterparts. On top of that, the $z\to 1$ region
is known to receive significant contributions from non-perturbative
physics, which has no analogue in QED. Finally, and most crucially,
while in QED $\Daem>0$, in QCD \mbox{$\Delta\as<0$}, which implies
that heavy-quark perturbative fragmentation functions are asymptotically 
strongly suppressed, rather than enhanced as for PDFs in QED. In conclusion,
the QCD case poses no problems, and the only potentially unpleasant fact
is that the strong suppression at $z\to 1$ of the NLL solution has no
LL counterpart. This has no phenomenological consequences.

\section{Singlet-photon large-$z$ solutions\label{sec:Sgaasy}}
In this section, we study the large-$z$ behaviour of the singlet
and photon NLL PDFs. Given the results of sect.~\ref{sec:NSasy},
we shall limit ourselves to working in the $\Delta_1$ framework.

The procedure we shall adopt follows closely that of appendix~B of 
ref.~\cite{Bertone:2019hks}. Because of this, we shall only sketch
it, and refer as much as possible to the equations of that paper.
While those equations complement the technical information given
here, they are not necessary for a general understanding of the strategy
used to arrive at the sought solutions. Such a strategy can be summarised 
as follows:
\begin{enumerate}
\item Write the evolution kernels in terms of a double series in $\aem$ 
and $1/N$.
\item Find the leading-$N$ solution in the form of a Magnus 
expansion~\cite{Magnus:1954,Magnus:rev2008}, which simplifies significantly 
owing to the kernels being diagonal at this order in $N$.
\item Treat the $1/N$ contributions as perturbations to the leading-$N$ 
evolution.
\item The solution is finally achieved in a standard way except for the fact
that it requires, in order for the inverse Mellin transforms to be computed 
analytically, a linearisation of the dependence of $\aem$ on the variable $t$.
\end{enumerate}
We start by recalling that by keeping only the divergent and constant 
terms of the Mellin transforms for $N\to\infty$ one fails to obtain
a good approximation of the actual $z\to 1$ behaviour of the photon
PDF. In $\MSb$, this is seen to stem from the fact that the electron-PDF LO 
initial condition is a Dirac delta that, combined with the relevant
off-diagonal term in the Altarelli-Parisi kernel, gives a contribution
of the same order as that due to the diagonal kernel element, times
the photon-PDF initial condition. This argument is independent of the 
factorisation scheme, but one needs to verify whether in a different
scheme, such as $\Delta$, contributions not present in an $\MSb$ computation
will significantly modify its conclusions. In view of this, the 
factorisation-scheme  matrix that we shall employ will have the
following form:
\beq
\Kmat_N^{(\Delta)}=\left(
\begin{array}{cccc}
K_{{\rm\sss S},N}^{(\Delta)} & 0 \\
K_{\gamma,N}^{(\Delta)} & 0 \\
\end{array}
\right)\,,
\label{JSgamat}
\eeq
which is eq.~(\ref{KmatDel}) written in the two-dimensional singlet-photon
sector, and where the functions $K_{{\rm\sss S},N}^{(\Delta)}$ and 
$K_{\gamma,N}^{(\Delta)}$ are the Mellin transforms of eqs.~(\ref{Kdelz}) 
and~(\ref{Kgadelz}), computed in the large-$N$ limit by including the 
leading terms suppressed by powers of $1/N$, namely:
\beqn
K_{{\rm\sss S},N}^{(\Delta)}\equiv K_N^{(\Delta)}
&\Melleq&
2\Big(\log^2\bN-\log\bN+C\Big)
+\frac{2}{N}\Big(\log\bN-\frac{3}{2}\Big)
\label{KNasy2}
\\*&\equiv&
K_{{\rm\sss S},N}^{[1,0]}+\frac{1}{N}\,K_{{\rm\sss S},N}^{[1,1]}\,,
\label{KNasy2bis}
\\*
K_{\gamma,N}^{(\Delta)}\equiv K_{\gamma e,N}^{(\Delta)}
&\Melleq&\frac{1}{N}\,.
\label{KgaNasy}
\eeqn
Clearly, eq.~(\ref{KNasy2}) extends eq.~(\ref{KNasy}) by adding to it
the relevant subleading terms. As was the case for the non-singlet
scheme-changing kernel, the matrix of eq.~(\ref{JSgamat}) is such that
the inverse operator of eq.~(\ref{invIpJ}) exists everywhere.

The evolution equation for the evolution operator, eq.~(\ref{EKopev}),
is re-written by introducing an evolution kernel as was done
in \ePDFteq{B.8}, namely:
\beq
\frac{\partial\Eop_N^{(\Delta_1)}(t)}{\partial t}=
\Mmat_N(t)\,\Eop_N^{(\Delta_1)}(t)\,,
\;\;\;\;\;\;\;\;
\Mmat_N(t)=\Mmat_N^{(A)}(t)+\Mmat_N^{(B)}(t)\,,
\eeq
with:
\beqn
\Mmat_N^{(A)}(t)&=&b_0\aem(\mu)\,\Kmat_N^{(\Delta)}
\left(I+\frac{\aem(\mu)}{2\pi}\,\Kmat_N^{(\Delta)}\right)^{-1}\,,
\label{matMNAdef}
\\
\Mmat_N^{(B)}(t)&=&
\left(I+\frac{\aem(\mu)}{2\pi}\,\Kmat_N^{(\Delta)}\right)
\left[\APmat_{{\rm\sss S},N}^{[0]}+\frac{\aem(\mu)}{2\pi}\left(
\APmat_{{\rm\sss S},N}^{[1]}-
\frac{2\pi b_1}{b_0}\,\APmat_{{\rm\sss S},N}^{[0]}\right)\right]
\nonumber\\*&&\phantom{aaaa}\times
\left(I+\frac{\aem(\mu)}{2\pi}\,\Kmat_N^{(\Delta)}\right)^{-1}\,,
\label{matMNBdef}
\eeqn
and where the Altarelli-Parisi kernels have been written as
in \ePDFteq{B.1} and \ePDFteq{B.2}:
\beqn
\APmat_{{\rm\sss S},N}&=&\APmat_{{\rm\sss S},N}^{[0]}+
\frac{\aem(\mu)}{2\pi}\APmat_{{\rm\sss S},N}^{[1]}+\ord(\aem^2)
\\&\equiv&
\left(\APmat_{{\rm\sss S},N}^{[0,0]}+
\frac{1}{N}\APmat_{{\rm\sss S},N}^{[0,1]}+\ord\left(N^{-2}\right)\right)
\nonumber\\*&&\phantom{aaa}
+\frac{\aem(\mu)}{2\pi}
\left(\APmat_{{\rm\sss S},N}^{[1,0]}+
\frac{1}{N}\APmat_{{\rm\sss S},N}^{[1,1]}+\ord\left(N^{-2}\right)\right)
+\ord(\aem^2)\,.
\label{APNSgsubN}
\eeqn
By construction, in $\MSb$ (i.e.~with $\Kmat=0$), $\Mmat_N(t)$ coincides
with that of \ePDFteq{B.7}. The forms in eqs.~(\ref{matMNAdef})
and~(\ref{matMNBdef}) stem from the observation that, as far as
the scheme-changing terms are concerned, an expansion in $\aem$
must {\em not} be carried out. Conversely, the contributions directly
related to the Altarelli-Parisi kernels can be safely expanded.

In keeping with eq.~(\ref{APNSgsubN}), we introduce
the leading and the first subleading terms of the kernels defined
in eqs.~(\ref{matMNAdef}) and~(\ref{matMNBdef}):
\beqn
\Mmat_N^{(A)}&=&\Mmat_N^{(A,0)}+\frac{1}{N}\,\Mmat_N^{(A,1)}+
\ord\left(N^{-2}\right)\,,
\label{matMNAex}
\\
\Mmat_N^{(B)}&=&\Mmat_N^{(B,0)}+\frac{1}{N}\,\Mmat_N^{(B,1)}+
\ord\left(N^{-2}\right)\,.
\label{matMNBex}
\eeqn
The pure leading-$N$ evolution will therefore be controlled
solely by the kernel:
\beq
\Mmat_N^{(0)}=\Mmat_N^{(A,0)}+\Mmat_N^{(B,0)}\,,
\label{Mmat0def}
\eeq
while both eq.~(\ref{Mmat0def}) and the following kernel:
\beq
\Mmat_N^{(1)}=\Mmat_N^{(A,1)}+\Mmat_N^{(B,1)}
\label{Mmat1def}
\eeq
will be relevant to computing the subleading corrections to the evolution.
With a direct calculation we obtain:
\beqn
\Mmat_N^{(A,0)}&=&
\left(
\begin{array}{cc}
\frac{\aem(\mu)b_0 K_{{\rm\sss S},N}^{[1,0]}}
{1+\frac{\aem(\mu)}{2\pi}K_{{\rm\sss S},N}^{[1,0]}} & 0 \\
0 & 0 \\
\end{array}
\right),
\label{matMNA0res}
\\
\Mmat_N^{(B,0)}&=&\APmat_{{\rm\sss S},N}^{[0,0]}+\frac{\aem(\mu)}{2\pi}
\left(\APmat_{{\rm\sss S},N}^{[1,0]}-
\frac{2\pi b_1}{b_0}\,\APmat_{{\rm\sss S},N}^{[0,0]}\right)\,,
\label{matMNB0res}
\eeqn
with eq.~(\ref{matMNB0res}) coinciding with \ePDFteq{B.12}.
Note that eq.~(\ref{matMNB0res}) is non trivial, because it is equivalent 
to saying that, at this order in $N$, the left and right multiplications 
by the scheme-changing matrices in eq.~(\ref{matMNBdef}) do not give
any extra contributions w.r.t.~the $\MSb$ result (indeed, this
conclusion is not valid starting from the first subleading order in $N$). 
Furthermore, owing to the presence of inverse matrices in the evolution
kernels, one may suspect that terms suppressed by higher powers of
$1/N$ in the Altarelli-Parisi and scheme-changing kernels (which have
been neglected here) could result, at the level of the evolution
kernels, in contributions of the same orders as those kept (i.e., $1/N$); 
we have explicitly verified that this is not the case. Following 
ref.~\cite{Bertone:2019hks}, the solution for the leading-$N$ 
evolution operator, which obeys:
\beq
\frac{\partial\Eop_N^{(\Delta_1,0)}(t)}{\partial t}=
\Mmat_N^{(0)}(t)\,\Eop_N^{(\Delta_1,0)}(t)\,,
\eeq
can be represented by means of the Magnus expansion~\cite{Magnus:1954,
Magnus:rev2008} (see \ePDFteq{B.8}). Since both of the matrices in 
eqs.~(\ref{matMNA0res}) and~(\ref{matMNB0res}) are diagonal,
so is their sum, and therefore only the first $\Omega$ term of the
Magnus series is non-zero. Furthermore, the exponent of a diagonal
matrix is the matrix whose elements are the exponents of the elements
of that matrix. This implies that we can write the sought solution
as follows:
\beq
\Eop_N^{(\Delta_1,0)}(t)=
\left(
\begin{array}{cc}
E_{\Sigma\Sigma,N}^{(A,0)} & 0\\
0 & 1 \\
\end{array}
\right)
\left(
\begin{array}{cc}
E_{\Sigma\Sigma,N}^{(B,0)} & 0\\
0 & E_{\gamma\gamma,N}^{(B,0)} \\
\end{array}
\right),
\label{Eop0}
\eeq
where:
\beqn
E_{\Sigma\Sigma,N}^{(A,0)}&=&
\frac{1+\frac{\aem(\mu)}{2\pi}K_{{\rm\sss S},N}^{[1,0]}}
{1+\frac{\aem(\mu_0)}{2\pi}K_{{\rm\sss S},N}^{[1,0]}}\,,
\label{EopA0SS}
\\
E_{\Sigma\Sigma,N}^{(B,0)}&=&\exp\left[-\xi_1\log\bN+\hat{\xi}_1\right],
\label{EopB0SS}
\\
E_{\gamma\gamma,N}^{(B,0)}&=&\exp\left[-\frac{2\NF}{3}t-
\frac{\aemmu-\aemz}{4\pi^2 b_0}\NF\Big(1-\frac{4\pi b_1}{3b_0}\Big)\right]
\label{EopB0gg}
\eeqn
with eqs.~(\ref{EopB0SS}) and~(\ref{EopB0gg}) coinciding with \ePDFteq{B.14}
and \ePDFteq{B.15}, respectively.
As it happens in $\MSb$, at the leading $N$ the singlet and photon
solutions factorise. From eqs.~(\ref{EopA0SS}) and~(\ref{EopB0SS}) we see 
that the former is identical to the one derived in sect.~\ref{sec:asyD1},
i.e.~eq.~(\ref{NLLsol8Del}). 
As far as the latter is concerned, eq.~(\ref{EopB0gg}) allows one to
make use of \ePDFteq{5.72}, that is:
\beq
M^{-1}\big[E_{\gamma\gamma,N}^{(B,0)}\big]\Melleq
\frac{\aem(\mu_0)}{\aem(\mu)}\,\delta(1-z)\,.
\label{Eggsol}
\eeq
Therefore, at this order in $N$ the photon PDFs is:
\beq
\Gamma_\gamma^{(\Delta_1)}(z,\mu)\stackrel{z\to 1}{\longrightarrow}
\frac{\aem(\mu_0)}{\aem(\mu)}\,\Gamma_\gamma^{(\Delta_1)}(z,\mu_0)\,.
\eeq
From eqs.~(\ref{Ggesol2}) and~(\ref{Kgadelz}) we thus obtain:
\beq
\Gamma_\gamma^{(\Delta_1)}(z,\mu)\stackrel{z\to 1}{\longrightarrow}
\frac{1}{2\pi}\,\frac{\aem^2(\mu_0)}{\aem(\mu)}\,\frac{1+(1-z)^2}{z}\,L_0\,,
\label{gaNLLsol3runDel}
\eeq
which is the analogue of \ePDFteq{5.73}. As is the case for the latter
equation, eq.~(\ref{gaNLLsol3runDel}) tends to a constant at $z\to 1$; 
but, at variance with what happens in $\MSb$, the leading-$N$ result
of the photon PDF in the $\Delta_1$ scheme is identically equal to
zero when $\mu_0=m$.

We now turn to considering the first subleading contribution, that
stems from the second terms on the r.h.s.~of eqs.~(\ref{matMNAex})
and~(\ref{matMNBex}), and is treated as a perturbation to the leading-$N$ 
solution. By proceeding as was done in ref.~\cite{Bertone:2019hks},
the evolution operator that includes both the leading and the first
subleading contributions is, according to \ePDFteq{B.30},
\beq
\Eop_N^{(\Delta_1)}(t)=\Eop_N^{(\Delta_1,0,L)}(t)
\left(I+\int_0^t dt_1\widehat{\Mmat}_N^{(1,L)}(t_1)\right)\,,
\label{EopNL}
\eeq
where, from \ePDFteq{B.19} and \ePDFteq{B.21}:
\beq
\widehat{\Mmat}_N^{(1,L)}(t)=\frac{1}{N}
\left(\Eop_N^{(\Delta_1,0,L)}(t)\right)^{-1}
\left[\Mmat_N^{(A,1)}(t)+\Mmat_N^{(B,1)}(t)\right]
\Eop_N^{(\Delta_1,0,L)}(t)\,.
\label{M1Nevol}
\eeq
Here, the linearised version of the leading-$N$ evolution operator
has been employed, which stems from eq.~(\ref{Eop0}) and reads as follows:
\beq
\Eop_N^{(\Delta_1,0,L)}(t)=
\left(
\begin{array}{cc}
E_{\Sigma\Sigma,N}^{(A,0)} & 0\\
0 & 1 \\
\end{array}
\right)
\left(
\begin{array}{cc}
E_{\Sigma\Sigma,N}^{(B,0,L)} & 0\\
0 & E_{\gamma\gamma,N}^{(B,0,L)} \\
\end{array}
\right),
\label{Eop0L}
\eeq
with:
\beqn
E_{\Sigma\Sigma,N}^{(B,0,L)}&=&\exp\left[\left(-\xi_{1,0}\log\bN
+\hat{\xi}_{1,0}\right)t\,\right]\,,
\label{EopB0LSS}
\\
E_{\gamma\gamma,N}^{(B,0,L)}&=&
\exp\left[-\left(\frac{2\NF}{3}+\chi_{1,0}\right)t\,\right]\,,
\label{EopB0Lgg}
\eeqn
with eqs.~(\ref{EopB0LSS}) and~(\ref{EopB0Lgg}) coinciding with \ePDFteq{B.23}
and \ePDFteq{B.24}, respectively, and where according to \ePDFteq{B.25}
the linearised expressions of the parameters $\xi_1$ and $\hat{\xi}_1$ 
of eqs.~(\ref{xiR1def}) and~(\ref{chi1Rdef}) have been employed:
\beq
\xi_1=\xi_{1,0}\,t+\ord(t^2)\,,
\;\;\;\;\;\;\;\;\;\;
\hat{\xi}_1=\hat{\xi}_{1,0}\,t+\ord(t^2)\,.
\label{hx10def}
\eeq
By means of an explicit computation one obtains:
\beqn
\xi_{1,0}&=&2\left[1-\frac{\aemz}{\pi}\left(\frac{5}{9}\NF+\frac{\pi b_1}{b_0}
\right)\right],
\label{xi10def}
\\
\hat{\xi}_{1,0}&=&
\frac{3}{2}\left[1+\frac{\aemz}{\pi}\left(\frac{\lambda_1}{3}\,
-\frac{\pi b_1}{b_0}\right)\right].
\label{hxi10def}
\eeqn
Note that the leftmost matrix on the r.h.s.~of eq.~(\ref{Eop0L}) is
not affected by the linearisation procedure (it is already linear
in $\aem(\mu)$). The matrices $\Mmat_N^{(A,1)}$ 
and $\Mmat_N^{(B,1)}$ that appear in eq.~(\ref{M1Nevol}) are computed
directly from their definitions in eqs.~(\ref{matMNAdef}), (\ref{matMNBdef}),
(\ref{matMNAex}), and~(\ref{matMNBex}). By using the same notation
as in eq.~(\ref{APNSgsubN}) to distinguish the $\ord(\aem^0)$ and $\ord(\aem)$ 
contributions:
\beqn
\Mmat_N^{(A,1)}&=&\frac{\aem(\mu)}{2\pi}\,\Mmat_N^{(A,[1,1])}\,,
\\
\Mmat_N^{(B,1)}&=&\Mmat_N^{(B,[0,1])}+
\frac{\aem(\mu)}{2\pi}\,\Mmat_N^{(B,[1,1])}\,,
\eeqn
we have obtained:
\beqn
\Mmat_N^{(A,[1,1])}&=&
\left(
\begin{array}{cc}
\frac{2\pi b_0 K_{{\rm\sss S},N}^{[1,1]}}
{\left(1+\frac{\aem(\mu)}{2\pi}K_{{\rm\sss S},N}^{[1,0]}\right)^2} & 0 \\
\frac{2\pi b_0}
{1+\frac{\aem(\mu)}{2\pi}K_{{\rm\sss S},N}^{[1,0]}} & 0 \\
\end{array}
\right),
\label{matMNA11res}
\\
\Mmat_N^{(B,[0,1])}&=&
\left(
\begin{array}{cc}
-1 & 2\NF \\
\frac{1}{1+\frac{\aem(\mu)}{2\pi}K_{{\rm\sss S},N}^{[1,0]}} & 0 \\
\end{array}
\right),
\label{matMNB01res}
\\
\Mmat_N^{(B,[1,1])}&=&
\left(
\begin{array}{cc}
\Mmat_{11,N}^{(B,[1,1])} & 
\Mmat_{12,N}^{(B,[1,1])} \\
\Mmat_{21,N}^{(B,[1,1])} & 
0 \\
\end{array}
\right),
\label{matMNB11res}
\eeqn
where:
\beqn
\Mmat_{11,N}^{(B,[1,1])}&\Melleq&
-4\log\bN+\frac{27+22\NF}{9}+\frac{2\pi b_1}{b_0}\,,
\label{matMNB11res11}
\\
\Mmat_{12,N}^{(B,[1,1])}&\Melleq&
2\NF\left(3\log^2\bN-2\log\bN+\frac{3+\pi^2}{6}-\frac{2\pi b_1}{b_0}\right)\,,
\label{matMNB11res12}
\\
\Mmat_{21,N}^{(B,[1,1])}&\Melleq&
\frac{1}{1+\frac{\aem(\mu)}{2\pi}K_{{\rm\sss S},N}^{[1,0]}}
\Bigg[-\log^2\bN+\frac{9+4\NF}{3}\,\log\bN
\nonumber
\\*&&\phantom{\frac{1}{1+\frac{\aem(\mu)}{2\pi}K_{{\rm\sss S},N}^{[1,0]}}}
-\frac{52\NF+3(27+\pi^2)}{18}-\frac{2\pi b_1}{b_0}\Bigg]\,.
\label{matMNB11res21}
\eeqn
From eq.~(\ref{KNasy2}) we see that the $(1,1)$ element 
in eq.~(\ref{matMNA11res}) is suppressed by an extra power of $\log\bN$
w.r.t.~to $(2,1)$ element; thus, it will be discarded henceforth.

Given the matrix elements above, one can compute the evolution
operator of eq.~(\ref{EopNL}). As in the $\MSb$ case, we are interested 
here only in the photon PDF (since for the singlet the leading-$N$ solution 
stemming from eqs.~(\ref{EopA0SS}) and~(\ref{EopB0SS}) is sufficient to give
a good description of the actual large-$z$ behaviour), and therefore we
shall only consider \ePDFteq{B.31}:
\beq
\Gamma_\gamma^{(\Delta_1)}(z)\Melleq
M^{-1}\Big[\big(\Eop_N^{(\Delta_1)}(t)\big)_{\gamma\Sigma}\,
\Gamma_{{\rm\sss S},0,N}\Big]
+M^{-1}\Big[\big(\Eop_N^{(\Delta_1)}(t)\big)_{\gamma\gamma}\,
\Gamma_{\gamma,0,N}\Big].
\label{gaPDFoffd}
\eeq
The initial conditions are still parametrised as in \ePDFteq{B.32}
and \ePDFteq{B.33}, that is:
\beqn
\Gamma_{{\rm\sss S},0,N}&=&1+\frac{\aemz}{2\pi}
\left(\Fzero+\Fone\log\bN+\Ftwo\log^2\bN\right)+\ord\left(N^{-1}\right)\,,
\label{asyelectN}
\\
\Gamma_{\gamma,0,N}&=&\ord\left(N^{-1}\right)\,,
\label{asygammaN}
\eeqn
but in the $\Delta$ scheme we have (eq.~(\ref{inicDelta})):
\beqn
F_0&=&\frac{3}{2}L_0\,,
\label{F0def}
\\
F_1&=&-2L_0\,,
\label{F1def}
\\
F_2&=&0\,,
\label{F2def}
\eeqn
to be compared with \ePDFteq{B.34}--\ePDFteq{B.36}. By taking this into
account, eq.~(\ref{gaPDFoffd}) has the same functional form as \ePDFteq{B.74},
namely:
\beq
\Gamma_\gamma^{(\Delta_1)}(z)\stackrel{z\to 1}{\longrightarrow}
\exp\left[-\left(\frac{2\NF}{3}+\chi_{1,0}\right)t\right]
\sum_{j=1}^5 \Gamma_{\gamma,j}^{(\Delta_1)}(z)\,.
\label{gaNLLsol4run}
\eeq
The $j=5$ contribution to eq.~(\ref{gaNLLsol4run}) stems from the 
second term on the r.h.s.~of eq.~(\ref{gaPDFoffd}). From
eq.~(\ref{gaNLLsol3runDel}) we have, analogously to \ePDFteq{B.73}:
\beq
\Gamma_{\gamma,5}^{(\Delta_1)}(z)=\frac{\aem(\mu_0)}{2\pi}\,
\frac{1+(1-z)^2}{z}\,L_0\,.
\label{gaNLLsol3run2}
\eeq
We recall that the form of eq.~(\ref{gaNLLsol4run}) is suited to taking
the fixed-$\aem$ limit. In general, the prefactor of that equation is
equal to \mbox{$\aem(\mu_0)/\aem(\mu)$} (see \ePDFteq{B.79}),
whence eq.~(\ref{gaNLLsol3run2}) times that prefactor coincides with
eq.~(\ref{gaNLLsol3runDel}), as it should. As far as the $j\le 4$
contributions to eq.~(\ref{gaNLLsol4run}) are concerned, they all stem 
from the first term on the r.h.s.~of eq.~(\ref{gaPDFoffd}). They
are formally similar to their $\MSb$ counterparts, \ePDFteq{B.38},
but are significantly different from them in practice. They read as follows:
\beq
\Gamma_{\gamma,j}^{(\Delta_1)}(z)\Melleq M^{-1}\left[\frac{1}{N}
\bN^{-\kappa_j}\frac{\sum_{i=0}^3 x_i^{(j)}\log^i\bN}
{\sum_{i=0}^3 y_i^{(j)}\log^i\bN}\right]\,,
\;\;\;\;\;\;\;\;\;\;j=1,2,3,4\,.
\label{gaPDFj14res}
\eeq
Thus, the largest power of $\log\bN$ in both the numerator and the
denominator of eq.~(\ref{gaPDFj14res}) is equal to $3$ (while it
is equal to $4$ and to $1$, respectively, in $\MSb$). This is 
true when $\aem\ne 0$. In the $\aem\to 0$ limit, i.e.~at the LL,
the $x_i^{(j)}$ and $y_i^{(j)}$ coefficients are such that the
numerator of eq.~(\ref{gaPDFj14res}) is $N$-independent, while
the denominator is of $\ord(\log\bN)$. Explicitly, one finds that
at $\aem\to 0$ eq.~(\ref{gaNLLsol4run}) coincides with \ePDFteq{B.88},
i.e.~with the $\MSb$ result; this is indeed what must happen, since
at the LL any scheme dependence disappears. At the NLL, the behaviour
of the numerator and denominator of eq.~(\ref{gaPDFj14res}) implies
that the photon PDF is at most a constant at $z\to 1$. We can easily 
compute such a constant; terms that vanish as some inverse power of
\mbox{$\log(1-z)$} would require a lengthy calculation (much more so
than in the case of $\MSb$), and this does not appear to be worth the
effort. We obtain what follows:
\beqn
\Gamma_\gamma^{(\Delta_1)}(z,\mu)&\stackrel{z\to 1}{\longrightarrow}&
\frac{1}{2\pi}\,\frac{\aem^2(\mu_0)}{\aem(\mu)}\,\frac{1+(1-z)^2}{z}\,L_0
+\frac{1}{2\pi\xi_{1,0}}\,\frac{\aem^2(\mu_0)}{\aem(\mu)}\,L_0
\nonumber\\*&&
-\frac{\aem(\mu)}{2\pi\xi_{1,0}}\,
\frac{e^{-\gE\xi_{1,0}t}e^{\hat{\xi}_{1,0}t}}
{\Gamma\left(1+\xi_{1,0}t\right)}\,(1-z)^{\xi_{1,0}t}\,L_0
\label{gaNLLsol4runQED}
\\*&\simeq&
\frac{1}{2\pi}\,\frac{\aem^2(\mu_0)}{\aem(\mu)}\,\frac{1+(1-z)^2}{z}\,L_0
+\frac{1}{2\pi\xi_1}\,\frac{t\,\aem^2(\mu_0)}{\aem(\mu)}\,L_0
\nonumber\\*&&
-\frac{t\,\aem(\mu)}{2\pi\xi_1}\,
\frac{e^{-\gE\xi_1}e^{\hat{\xi}_1}}
{\Gamma\left(1+\xi_1\right)}\,(1-z)^{\xi_1}\,L_0\,.
\label{gaNLLsol4runQED2}
\eeqn
Eq.~(\ref{gaNLLsol4runQED2}) is obtained from eq.~(\ref{gaNLLsol4runQED})
by undoing, at the purely parametric level, the linearisation introduced
to deal with the subleading-$N$ evolution operator; this stems from
eq.~(\ref{hx10def}). 
While the undoing of the linearisation is not justified mathematically,
it seems to be fairly harmless, and if so eq.~(\ref{gaNLLsol4runQED2})
should be considered as our best prediction for the large-$z$ photon PDF.

The first term on the r.h.s.~of eq.~(\ref{gaNLLsol4runQED2}) stems
from the leading-$N$ contribution, whereas the second and third terms
stem from the subleading contribution. In the strict $z\to 1$ limit,
where all terms that vanish need to be ignored, the first and the second
terms are equal to non-zero constants of the same order, while the third
term vanishes. Thus, by neglecting higher-order terms (in $\aem$) 
originating from the expansion of $\xi_1$, we have:
\beq
\Gamma_\gamma^{(\Delta_1)}(z,\mu)\;\stackrel{z\to 1}{\longrightarrow}\;
\frac{1}{2\pi}\,\frac{\aem^2(\mu_0)}{\aem(\mu)}\,L_0
+\frac{1}{4\pi}\,\frac{\aem^2(\mu_0)}{\aem(\mu)}\,L_0+0
=\frac{3}{4\pi}\,\frac{\aem^2(\mu_0)}{\aem(\mu)}\,L_0\,.
\eeq
This has to be compared with its analogue in $\MSb$, eq.~(\ref{ePDFgaasy}),
which is divergent at $z\to 1$, and is so that such a divergence
is entirely driven by the subleading-$N$ solution of the evolution
equation. While this does not happen in the $\Delta$ scheme, also in
this case it is true that subleading-$N$ contributions cannot be
neglected for the computation of the photon PDF to be reliable in
the $z\to 1$ region.

Finally, note that eq.~(\ref{gaNLLsol4runQED2}) vanishes identically
when $L_0=0$. This is a consequence of the fact that, in the $\Delta$
scheme, when $L_0=0$ all initial conditions are equal to zero, except for
the $\ord(\aem^0)$ one of the electron (which is equal to $\delta(1-z)$).
In turn, this implies that the largest power of $\log\bN$ in the
numerator of eq.~(\ref{gaPDFj14res}) is equal to $2$, while that in
the denominator is equal to $3$. Therefore, at $L_0=0$ the photon
PDF is in general non-zero for a generic $z$, but vanishes at least
as fast as \mbox{$\log^{-1}(1-z)$} at $z\to 1$.

\section{Conclusions\label{sec:concl}}
In this paper, we have studied analytically the behaviour of the
NLL-accurate electron and photon PDFs of the electron in the $z\to 1$ 
region, by considering a DIS-like factorisation scheme, called $\Delta$. 
This region gives the dominant contribution to $\epem$ cross sections, 
and is therefore important, in view of the precision targets for theoretical 
calculations to be achieved in the context of the physics programs of future 
lepton colliders, that the systematics associated with the factorisation-scheme 
dependence be carefully assessed. This can be done by comparing predictions 
for any given process obtained with PDFs evolved with different 
factorisation schemes, such as those derived here and their $\MSb$ analogues, 
previously computed at the NLL accuracy in ref.~\cite{Bertone:2019hks}.

The main results of this work are given in eqs.~(\ref{NLLsol8Del})
and~(\ref{NLLsol9Del}) (for both the singlet and the non-singlet),
and eq.~(\ref{gaNLLsol4runQED2}) (for the photon).
We have shown that, while the definition of the $\Delta$ scheme aims
to obtain NLL-accurate PDFs which are as close as possible to their
LL counterparts, this happens only if suitably-defined initial 
conditions are evolved by including the effects of the running
of $\aem$ to all orders. A truncation of running-$\aem$ effects
to some order in $\aem$ in the evolution equation causes the 
electron PDF to develop a non-integrable singularity at $z=1$.
This is a result which is peculiar to QED, and its QCD analogue
(the fragmentation function of a heavy quark) has no comparable issue.

When the running of $\aem$ is properly taken into account, the NLL PDFs
in the $\Delta$ scheme have the same $z\to 1$ functional forms as
their LL analogues. This is not the case for the $\MSb$ scheme, whose
PDFs feature additional \mbox{$\log(1-z)$} terms w.r.t.~those possibly
present at the LL. While this is a positive characteristics from a 
theoretical viewpoint, in view of the fact that PDFs are non-physical 
quantities its phenomenological implications can be assessed only after 
computing observable cross sections.

\section*{Acknowledgments}
I am indebted to V.~Bertone, M.~Cacciari, G.~Stagnitto, M.~Zaro, 
and X.~Zhao, with whom I collaborate on topics related to those studied
here, for the many discussions we have had during the course of this work.
I also thank Marco Bonvini, Carlo Carloni Calame, and Fulvio Piccinini 
for having provided me with useful information.

\appendix
\section{Some useful formulae for inverse Mellin transforms\label{sec:Mell}}
We are interested in computing the following inverse Mellin transform:
\beq
M^{-1}\!\left[N^{-\xi}\log^q\bN\right]
\label{invMlq}
\eeq
for any positive real number $\xi$ and positive integer $q$. In order 
to do so, we introduce an arbitrary real parameter $\beta$, so that:
\beq
e^{\beta\log\bN}=\sum_{q=0}^\infty\frac{\beta^q}{q!}\log^q\bN\,.
\eeq
Therefore, by assuming uniform convergence:
\beq
M^{-1}\!\left[N^{-\xi}\log^q\bN\right]=
\frac{\partial^q}{\partial\beta^q}\left.
M^{-1}\!\left[N^{-\xi}e^{\beta\log\bN}\right]\right|_{\beta=0}\,.
\label{invMlogqN2}
\eeq
Since
\beq
M\left[(1-z)^{-1+\rho}\right]\,\Melleq\,
\Gamma(\rho)N^{-\rho}
\label{Mmoomz}
\eeq
for any positive $\rho$, then:
\beqn
&&M^{-1}\!\left[N^{-\xi}e^{\beta\log\bN}\right]\,\Melleq\,
\frac{e^{\gE\beta}}{\Gamma(\xi-\beta)}\,(1-z)^{-1+\xi-\beta}
\label{invMexp1}
\\*&&\phantom{aaaaaaa}
=\exp\Big[\gE\beta-(1-\xi+\beta)\log(1-z)-\log\Gamma(\xi-\beta)\Big]
\,.\phantom{aaa}
\label{invMexp2}
\eeqn
The expression in eq.~(\ref{invMexp2}) is well suited to the computation
of the derivatives of interest. In fact, one can apply the Fa\`a di Bruno 
formula for composite derivatives, namely:
\beq
\frac{d^q f(g(\beta))}{d\beta^q}=
\sum_{i=1}^q f^{(i)}(g(\beta))\,B_{q,i}\big(g^{(1)}(\beta),g^{(2)}(\beta),
\ldots g^{(q-i+1)}(\beta)\big)\,,
\label{FaadB}
\eeq
valid for any two functions $f$ and $g$. In our case, from 
eq.~(\ref{invMexp2}) we have:
\beqn
f(z)&=&e^z\,,
\label{eFaa1}
\\
g(\beta)&=&
\gE\beta-(1-\xi+\beta)\log(1-z)-\log\Gamma(\xi-\beta)\,.
\label{eFaa2}
\eeqn
In eq.~(\ref{FaadB}), by $B_{q,i}$ we have denoted the {\em incomplete}
exponential Bell polynomials. In fact, given eq.~(\ref{eFaa1}),
all of the derivatives of the function $f$ in eq.~(\ref{FaadB}) are
identical to the function itself, and can therefore be factored out
from the summation. In this way, the {\em complete} exponential 
Bell polynomials:
\beq
B_q(x_1,\ldots x_q)=\sum_{i=1}^q B_{q,i}\big(x_1,\ldots x_{q-i+1}\big)
\eeq
emerge naturally in eq.~(\ref{FaadB}). By putting all this together, we
therefore obtain:
\beq
M^{-1}\!\left[N^{-\xi}\log^q\bN\right]\,\Melleq\,
\frac{(1-z)^{-1+\xi}}{\Gamma(\xi)}\,
B_q\big(\Phi_q(\xi)\big)\,,
\label{invMlogqN3}
\eeq
where we have introduced the  $q$-dimensional vector $\Phi_q$, whose
components, according to eq.~(\ref{FaadB}), are:
\beq
\Phi_q^{[i]}(\xi)=
\left.\frac{d^i g(\beta)}{d\beta^i}\right|_{\beta=0}.
\eeq
With the explicit form of eq.~(\ref{eFaa2}), this leads to:
\beqn
\Phi_q(\xi)&=&\Psi_q(\xi)+\big(-\log(1-z)\,,\vec{0}_{q-1}\big)\,,
\label{Phidef}
\\*
\Psi_q(\xi)&=&\left(\gE+\psi_0(\xi),-\psi_1(\xi),\ldots,
(-)^{q-1}\psi_{q-1}(\xi)\right)\,,
\label{Psivec}
\eeqn
where $\psi_k(\xi)$ are the polygamma functions:
\beq
\psi_k(\xi)=\frac{d^{k+1}}{d\xi^{k+1}}\,\log\Gamma(\xi)\,.
\eeq
We can now exploit the following sum rule of the complete exponential
Bell polynomials:
\beq
B_q(x_1+y_1,\ldots x_q+y_q)=\sum_{i=0}^q \binomial{q}{i}
B_{q-i}(x_1,\ldots x_{q-i})\,B_i(y_1,\ldots y_i)\,,
\label{BqvsBi}
\eeq
and the identity:
\beq
B_i(y_1,0,\ldots 0)=y_1^i\,.
\label{Bizero}
\eeq
This leads to an explicit expression for the inverse Mellin transform
of eq.~(\ref{invMlq}), namely\footnote{For results of similar kind as
that of eq.~(\ref{invMlogqN}), see e.g.~refs.~\cite{Blumlein:1998if,
Bonvini:2012sh,Almasy:2015dyv}.}:
\beq
M^{-1}\!\left[N^{-\xi}\log^q\bN\right]\,\Melleq\,
\frac{\xi(1-z)^{-1+\xi}}{\Gamma(1+\xi)}
\sum_{i=0}^q (-)^{q-i}\binomial{q}{i}d_i(\xi)\,\log^{q-i}(1-z)\,,
\label{invMlogqN}
\eeq
where:
\beq
d_k(\xi)=B_k\big(\Psi_k(\xi)\big)\,.
\label{dkBell}
\eeq
The derivation above, and the role that the Bell polynomials play in the
Fa\`a di Bruno formula, suggest an alternative form for the coefficients
$d_k$. Indeed, one can verify by employing that formula that the quantity:
\beq
{\cal G}_d(\xi,\delta)=
\frac{e^{-\gE(\xi-\delta)}}{\Gamma(\xi-\delta)}\,
\label{dffgenfun}
\eeq
can be used as the generating functional of the $d_k$ coefficients,
namely:
\beq
d_k(\xi)=\frac{1}{{\cal G}_d(\xi,0)}
\left.\frac{\partial^k{\cal G}_d(\xi,\delta)}{\partial\delta^k}
\right|_{\delta=0}\,.
\label{dffbygen}
\eeq
The results of eqs.~(\ref{dkBell}) and~(\ref{dffbygen}) coincide. 
We mention the fact that the functions $A(\xi)$ and $B(\xi)$,
introduced in ref.~\cite{Bertone:2019hks} and that appear in
eq.~(\ref{NLLsol3run}), are such that:
\beqn
d_1(\xi)&=&-A(\xi)\,,
\label{dores}
\\
d_2(\xi)&=&2B(\xi)-\frac{\pi^2}{6}\,.
\label{dtres}
\eeqn
The form of eq.~(\ref{dkBell}) allows one to find relationships 
among the $d_k$ coefficients. Explicitly, by using the following
recursion relation for the complete Bell polynomials:
\beq
B_k(x_1,\ldots x_k)=\sum_{i=1}^k\binomial{k-1}{i-1}
x_i\,B_{k-i}(x_1,\ldots x_{k-i})\,,
\label{cBrec}
\eeq
we obtain:
\beq
d_k(\xi)=\sum_{i=1}^k\binomial{k-1}{i-1}
\Psi_k^{[i]}(\xi)\,d_{k-i}(\xi)\,.
\label{dkrec}
\eeq
In fact, eq.~(\ref{dkrec}) can be used to compute the $d_k$ coefficients
with increasing values of $k$, starting from $k=0$ and using the fact that:
\beq
B_0(x)=1\;\;\;\;\Longrightarrow\;\;\;\;d_0(\xi)=1\,.
\label{dkzero}
\eeq
Furthermore, the expression of the $d_k(\xi)$ coefficients in terms
of the Bell polynomials allows one to compute the $\xi\to 0$ behaviour
of such coefficients, which is directly relevant to the electron PDFs
(where $\xi=\xi_1$, and $\xi_1\simeq 0.05$ for scales of 
$\ord(100~{\rm GeV})$). This is done as follows. Consider:
\beq
S=\sum_{k=0}^\infty d_k(\xi)\frac{t^k}{k!}\equiv
\sum_{k=0}^\infty B_k\big(\Psi_k(\xi)\big)\frac{t^k}{k!}=
\exp\left[\sum_{j=1}^\infty \Psi_\infty^{[j]}(\xi)\frac{t^j}{j!}\right]\,,
\label{Sdef}
\eeq
where the rightmost equality follows from the fact that the complete 
exponential Bell polynomials satisfy the generating-functional relation:
\beq
\exp\left(\sum_{j=1}^\infty x_j\,\frac{t^j}{j!}\right)=
\sum_{k=0}^\infty B_k(x_1,\ldots x_k)\frac{t^k}{k!}\,.
\label{BellGenF}
\eeq
By construction:
\beq
d_k(\xi)=\left.\frac{\partial^k S}{\partial t^k}\right|_{t=0}\,.
\label{dkfromS}
\eeq
We can now use the following representation of the polygamma functions:
\beq
\psi_k(\xi)=-\gE\delta_{k0}-\frac{(-)^k k!}{\xi^{k+1}}+
\sum_{i=1}^\infty\left(\frac{\delta_{k0}}{i}-\frac{(-)^k k!}{(i+\xi)^{k+1}}
\right)\,,
\label{psik}
\eeq
to write their small-$\xi$ behaviour as follows:
\beqn
\psi_k(\xi)&=&-\gE\delta_{k0}-\frac{(-)^k k!}{\xi^{k+1}}+
\sum_{i=1}^\infty\left(\frac{\delta_{k0}}{i}-\frac{(-)^k k!}{i^{k+1}}
\right)
\label{psiksxi}
\\*&&
+\xi\,(-)^k(k+1)!\sum_{i=1}^\infty\frac{1}{i^{k+2}}
-\frac{\xi^2}{2}(-)^k(k+2)!\sum_{i=1}^\infty\frac{1}{i^{k+3}}
+\ord(\xi^3)\,.
\nonumber
\eeqn
The series on the r.h.s.~of this equation can easily be summed,
leading to:
\beqn
\psi_k(\xi)&=&-\frac{(-)^k k!}{\xi^{k+1}}-\gE\delta_{k0}
-(-)^k k!\,\zeta_{k+1}\left(1-\delta_{k0}\right)
\label{psiksxi2}
\\*&&
+\xi\,(-)^k (k+1)!\,\zeta_{k+2}
-\frac{\xi^2}{2}(-)^k(k+2)!\,\zeta_{k+3}
+\ord(\xi^3)\,.
\nonumber
\eeqn
We can use this expression for evaluating the arguments of the Bell
polynomials in eq.~(\ref{Sdef}); according to eq.~(\ref{Psivec}):
\beq
\Psi_\infty^{[j]}(\xi)=\gE\delta_{j1}+(-)^{j-1}\psi_{j-1}(\xi)\,.
\eeq
With this, we obtain:
\beqn
\sum_{j=1}^\infty \Psi_\infty^{[j]}(\xi)\frac{t^j}{j!}&=&
\log\left(1-\frac{t}{\xi}\right)+\gE t-\log\Gamma(1-t)
\\*&&
-\big(\gE+\psi_0(1-t)\big)\xi+
\left(\frac{\pi^2}{12}-\half\psi_1(1-t)\right)\xi^2+
\ord(\xi^3)\,.
\nonumber
\eeqn
Therefore:
\beqn
S\!\!&=&\!\!\frac{\xi-t}{\xi\Gamma(1-t)}
\label{Sres}
\\*&&\times
\exp\left[\gE t -\big(\gE+\psi_0(1-t)\big)\xi+
\left(\frac{\pi^2}{12}-\half\psi_1(1-t)\right)\xi^2+
\ord(\xi^3)\right]\,.
\nonumber
\eeqn
In the computation of the derivatives of interest, eq.~(\ref{dkfromS}),
it is a simple matter of algebra to see that:
\beq
d_k(\xi)=-\frac{k}{\xi}\left.\frac{\partial^{k-1}}{\partial t^{k-1}}
\frac{e^{\gE t}}{\Gamma(1-t)}\right|_{t=0}
+\ord(\xi^0)\,.
\label{dkasy}
\eeq
This proves that all $d_k$ coefficients diverge at most as $1/\xi$ at 
small $\xi$, which is a remarkable fact in view of eq.~(\ref{psiksxi}),
that exhibits a much steeper $\xi\to 0$ behaviour. By using the Fa\`a di 
Bruno formula in eq.~(\ref{dkasy}) we arrive at:
\beq
d_k(\xi)=-\frac{k}{\xi}\,B_{k-1}\big(Z_{k-1}\big)+\ord(\xi^0)\,,
\label{dkasy2}
\eeq
having defined the $(k-1)$-dimensional vector:
\beq
Z_{k-1}=\big(0,\,-1!\,\zeta_2,\,-2!\,\zeta_3,
\ldots,\, -(k-2)!\,\zeta_{k-1}\big)\,.
\label{Zvec}
\eeq
We have verified up to $k=\ord(25)$ that the result of eq.~(\ref{dkasy2}) 
is in agreement with that obtained by Taylor-expanding in $\xi$ the complete 
result for $d_k(\xi)$ of either eq.~(\ref{dkBell}) or eq.~(\ref{dffbygen}); 
while computing the former is straightforward, the computation of the latter 
becomes involved for Mathematica, starting at about $k=10$.

Before concluding this section, we remark that eq.~(\ref{invMlogqN})
can be used to arrive at a minor improvement of the result for the $\MSb$ 
photon PDF w.r.t.~that obtained in ref.~\cite{Bertone:2019hks}. This 
is due to applying eq.~(\ref{invMlogqN}) in \ePDFteq{B.39} and 
\ePDFteq{B.40}\footnote{Note that in eq.~(\ref{invMlogqN}) there is a 
factor $N^{-\kappa}$, whereas \ePDFteq{B.39} and \ePDFteq{B.40} a 
factor $\bN^{-\kappa}$. Furthermore, eq.~(\ref{invMlogqN}) must be used 
after having performed the formal replacement \mbox{$\kappa\to 1+\kappa$}.},
which leads to the possibility of computing the functions in \ePDFteq{B.41}, 
\ePDFteq{B.43}, \ePDFteq{B.44}, and \ePDFteq{B.45} exactly in $\kappa$, rather 
than by means of  an $\ord(\kappa^2)$ expansion as in 
ref.~\cite{Bertone:2019hks}. We obtain:
\beqn
\invm_1(z;\kappa,\dencpar,\dendpar)\!&=&\!
\frac{e^{-\gE\kappa}(1-z)^\kappa}{\Gamma(1+\kappa)}\,
\nonumber\\*&&\times
\sum_{i=0}^\infty \frac{(-)^i\dencpar^i\,d_i(1+\kappa)}
{\left(\dendpar-\dencpar\log(1-z)\right)^{i+1}}\,,
\label{Mmoinv1res}
\\
\invm_{p+2}(z;\kappa,\dencpar,\dendpar)\!&=&\!
\frac{e^{-\gE\kappa}(1-z)^\kappa}{\Gamma(1+\kappa)}\,
\nonumber\\*&&\times
\sum_{i=0}^p(-)^{p-i}\binomial{p}{i}d_i(1+\kappa)\,\log^{p-i}(1-z)\,,
\;\;\;\;\;\;0\le p\le 3\,,
\phantom{aaaaaa}
\label{Mmoinvpp2res}
\eeqn
with the $d_i$ coefficients given in eq.~(\ref{dkBell}) or 
eq.~(\ref{dffbygen}). 

In practice, the parameter
$\kappa$ actually used in the computation of the $\MSb$ photon PDF
is either very small or equal to zero; thus, the numerical results 
of eqs.~(\ref{Mmoinv1res}) and~(\ref{Mmoinvpp2res}) do not differ either
significantly or at all from those of ref.~\cite{Bertone:2019hks}. Note 
that \ePDFteq{B.41} corresponds to limiting the summation on the r.h.s.~of 
eq.~(\ref{Mmoinv1res}) to $i\le 2$. Conversely, the infinite sum of
eq.~(\ref{Mmoinv1res}) can be formally expressed by means of a Borel
integral, thanks to eq.~(\ref{dffbygen}):
\beqn
\invm_1(z;\kappa,\dencpar,\dendpar)\!&=&\!
\frac{e^{-\gE\kappa}(1-z)^\kappa}{\Gamma(1+\kappa)}\,
\left(\Big(\dendpar-\dencpar\log(1-z)\right)\,
{\cal G}_d(1+\kappa,0)\Big)^{-1}
\nonumber\\*&&\times
\int_0^\infty dt\,e^{-t}\,
{\cal G}_d\left(1+\kappa,-\,\frac{t\,\dencpar}
{\dendpar-\dencpar\log(1-z)}\right)\,.
\label{Mmoinv1resBS}
\eeqn
In view of eq.~(\ref{dffgenfun}), the integral on the r.h.s.~of
eq.~(\ref{Mmoinv1resBS}) cannot be computed analytically, and therefore
this is a result of limited interest.

\phantomsection
\addcontentsline{toc}{section}{References}
\bibliographystyle{JHEP}
\bibliography{eepdfdel}

\end{document}